\documentclass[11pt]{article}
\usepackage[a4paper, total={6in, 8in}]{geometry}
\usepackage{epsfig,enumerate,verbatim}
\usepackage{amsmath}
\usepackage{amsfonts}
\usepackage{amssymb}
\usepackage{amstext}
\usepackage{amsthm}
\usepackage{dsfont}
\usepackage{graphicx}
\usepackage{subcaption}
\usepackage{mathrsfs}
\usepackage{cite}
\begin{document}        
\title{\bf{Deformed Gazeau-Klauder Schr$\ddot{\text{o}}$dinger-Cat States with Modified Commutation Relations}}
\author{C.L. Ching\footnote{Email: cheeleong\textunderscore ching@sutd.edu.sg (corresponding author)}, 
and W.K. Ng\footnote{Email: phynwk@nus.edu.sg}}

\date{3$^{\textrm{th}}$ Oct 2019}

\maketitle

\begin{center}
{$\ast$ Science and Math Cluster, Singapore University of Technology and Design (SUTD),\\}
{Upper Changi, Singapore 487372.\\}
{$\dagger$ Department of Physics, National University of Singapore (NUS),\\}
{Kent Ridge, Singapore 117551.}

\end{center}

\begin{abstract}
\noindent Generalized coherent states (GCs) under deformed quantum mechanics which exhibit intrinsic minimum length and maximum momentum have been well studied following Gazeau-Klauder approach in Refs.\cite{subirCS,deyfringCS,pedramCS,GCSmaxP}. In this paper, as an extension to the study of quantum deformation, we investigate the famous Schr$\ddot{\text{o}}$dinger cat states (SCs) under these two classes of quantum deformation. Following the concept of generalized Gazeau-Klauder Schr$\ddot{\text{o}}$dinger cat states (GKSCs) in\cite{dajka}, we construct the deformed-GKSCs for both phenomenological models that exhibit intrinsic minimum length and/or maximum momentum. All comparisons between minimum length and maximum momentum deformations are illustrated and plots are done in even and odd cat states since they are one of the most important classic statistical characteristics of SCs. Probability distribution and entropies are studied. In general, deformed cat states do not possess the original even and odd states statistical properties. Nonclassical properties of the deformed-GKSCs are explored in terms of Mandel $Q$ parameter, quadrature squeezing $(\Delta X_q)\cdot(\Delta Y_q)$ as well as Husimi quasi-probability distribution $\mathcal{Q}$. Some of these distinguishing quantum-gravitational features may possibly be realized qualitatively and even be measured quantitatively in future experiments with the advanced development in quantum atomic and optics technology.
\end{abstract}


\section{Introduction}

After almost more than half of a century of intense research activities to reconcile the two main pillars of modern physics, namely quantum mechanical theory which governs the subatomic world and general relativity which describes physics at the cosmological scale, up to date there is still no common consensus on the form of new gravitational effects that may show up in quantum system at high energy. Nevertheless, there are a few partially promising candidates for quantum theory of gravity such as Superstrings Theory\cite{polchinski,becker}, Loop Quantum Gravity (LQG)\cite{rovelli,ashtekar}, Noncommutative Geometry (NC)\cite{connes,douglas}, Renormalization Group (RG) flow/Asymptotic Safety\cite{reuter} approach and etc. All of these approaches predict some important generic  physical features such as the existence of a minimal resolution length scale $l_{min}$, probably taken value at about the order of Planck length $L_{p}=\sqrt{\frac{\hbar G}{c^3}}\approx 1.6 \times 10^{-35}$m\cite{adler}. Here $\hbar$ is the Planck constant, $G$ is the Newton gravitational constant and $c$ is the speed of light. From the phenomenological point of view, modified quantum commutation relations (MCRs) have been extensively studied as effective means of encoding potential gravitational or stringy/loopy effects; see Refs.\cite{snyder,kempf1,kempf2,laynam1,lewis,laynam2,quesne1,brau,harbach,dasvagenas,nozari,bouaziz,kober,pedram1} and the reviews\cite{sabine,tawfik} for a complete list of references. While most of the studied MCRs incorporate a minimum position uncertainty and usually lead to the concept of a minimal length scale thus consistent with quantum gravity phenomenology, there are others that exhibit a maximum momentum; see Refs.\cite{magueijo1,magueijo2,cortes,camelia,das1,das2,das3,pedram2,mignemi,jizba,maggiore,battisti,pedram3,chingspectra} as in doubly special relativity (DSR)\cite{camelia-dsr,magueijo-dsr} and Anti-Snyder model\cite{snyder,mignemi}. It has also been suggested that the consequent deformations of quantum mechanical spectra and relevant physics might be detectable in future low-energy experiments\cite{pikovski,marin,bawaj,gao,rossi,deyopto,howl}. In Ref.\cite{chingspectra}, we investigated in detail large classes of deformed quantum mechanics of the latter type and their implication in quantum optical systems\cite{GCSmaxP}. Current manuscript is an extension to these previous works. 

On the other hand, coherent states (Cs) and squeezed states (Ss) are very interesting quantum systems in nature due to their ability to exhibit the ``quantumness" ranging from almost classical\footnote{Coherent states saturate Heisenberg Uncertainty Principle (HUP) in vacuum state hence it is regarded as state closest to classical physics, while Ss showed highly nontrivial quantum randomness in their sub-Poisson distribution and anti-bunching behavior.} to highly nonclassical features such as quadrature squeezing below HUP boundary, sub-Poisonnian statistics, anti-bunching effects and quasi probability distribution; see Refs.\cite{klauder0,klauder1,glauber1,sudarshan1,gazeau1,klauderB,scully,mandel,walls,loudon,angelova}. These interesting features enable both coherent and squeezed states to be extremely useful in quantum information processing\cite{kok1,kok2}, optical communication and measurement\cite{yamamoto}, quantum metrology\cite{giovannetti}, quantum cryptography\cite{hillery} and etc. Particularly, squeezed states have been used in LIGO to increase the sensitivity of gravitational wave detectors since one can achieve lower noise in spatial quadrature\cite{anisimov,aasi}. Furthermore, recently there is an increase of interest to study generalized squeezed state in noncommutative and deformed space settings\cite{deyfringCS,GCSmaxP,deyq,deyfringR,deyhussin,deyzelaya,curado,berrada} in order to capture the nonconventional quantum gravitational effects\cite{howl} and also relations to quantum nonlocality\cite{yingwu}.   

Our manuscript is organized as follows: In Sect.(2), we review the formalism of generalized Heisenberg algebra (GHA)\cite{gha1,gha2,gha3} in constructing the so called Gazeau-Klauder's Coherent States (GKCS)\cite{gazeau2,klauder1,klauder2,klauder3} and further obtain their superposition as the Gazeau-Klauder Schr$\ddot{\text{o}}$dinger Cat States (GKSCs)\cite{dajka,deyfringR}. In Sect.(3), after reviewing the two important quantum gravity inspired phenomenological models which exhibit minimal length and/or maximum momentum scale, we explicitly construct the deformed-GKSCs. We study the probability distribution and entropies of the odd and even deformed-GKSCs in Sect.(4). The nonclassical behaviors of the cat states are further explored in terms of number and quadrature squeezing in Sect(5) as well as Husimi's distribution in Sect.(6). Finally we conclude in Sect.(7).        

\section{Gazeau-Klauder Coherent and Cat States}

Coherent states (Cs) were first studied by Schrodinger in 1926 in harmonic oscillator systems\cite{schrodinger} and later by Klauder and Glauber\cite{klauder0,klauder1,glauber1,sudarshan1,klauder2,klauder3,gazeau2}. Glauber obtained these states in the study of electromagnetic correlation function and realized the interesting feature that these states saturate the HUP, $\Delta x\Delta p= \hbar/2$. Thus, coherent states are considered as quantum states with the closest behavior to the classical system and have many applications in theoretical and mathematical physics\cite{csapp1,csapp2,csapp3}. 

In literature, there are two ways to construct the coherent states. First is through the Klauder\rq{}s approach by using the Fock\rq{}s representation of ladder algebra and second is the Perelomov-Gilmore\rq{}s approach\cite{gilm} based on group theoretic construction. In this paper, we follow Klauder\rq{}s approach and use the generalized Heisenberg algebra (GHA) given in\cite{gha1,gha2,gha3}. In this version of GHA\footnote{There are other versions of GHA. Back in early 1950, E. Wigner posed an intriguing question ``\textit{Do the equations of motion determine the quantum mechanical commutation relations?}". According to Wigner, equation of motion has a more immediate physical significance  than Heisenberg commutation relation $[x_i,p_j]=i\hbar\delta_{ij}$. He found as an answer a generalized quantum mechanical rule (deformation was introduced implicitly here) for the one-dimensional harmonic oscillator \cite{wigner,yang}. Wigner's idea was further explored and it leads to the new deformed quantum commutation relation generally called Wigner-Heisenberg Algebra (WHA). This algebra subsequently found many important and interesting physical applications related to quantum chromodynamics\cite{greenberg}, parastatistics\cite{plyushchay1,plyushchay2,plyushchay3}, anyons physics\cite{plyushchay4,plyushchay5} and supersysmmetry\cite{plyushchay6}. For recent application of WHA in the context of coherent state, Schrodinger cat states and quantum entanglement transfer, see Ref.\cite{deghani1,deghani2,dehdashti,deghani3}.}, the Hamiltonian $J_{0}$ which is related to the characteristic function $g(x)$ of the physical system, together with ladder operators ($A^{\dagger}$ being creation operator and $A=(A^{\dagger})^{\dagger}$ being the annihilation operator) play the role of the generators of the algebra,
\begin{eqnarray}
J_{0} A^{\dagger} &=& A^{\dagger} g(J_{0})\nonumber\\
A J_{0} &=& g(J_{0}) A\nonumber\\
\left[A^{\dagger},A\right] &=& J_{0} - g(J_{0})\ . \label{gha}
\end{eqnarray}    
We see that these operators form a closed algebra and $g(J_{0})$ is the analytic function of $J_{0}$ which is unique for each type of GHA. The Casimir of this algebra is 
\begin{eqnarray}
C &=& A^{\dagger}A- J_{0}=A A^{\dagger} - g(J_{0})\ .
\end{eqnarray}
The vacuum of the generator $J_{0}$ is defined by 
\begin{eqnarray}
J_{0}|0\rangle &=&\alpha_{0}|0\rangle \label{vacuum1}
\end{eqnarray}
where $\alpha_{0}$ is the energy eigen-value of the vacuum state $|0\rangle$. Also, the vacuum is annihilated by the operator $A$, i.e. $A |0\rangle=0$. Consider a general eigen-ket of $J_{0}$, denoted by $|m\rangle$, the generators satisfy the following,
\begin{eqnarray}
J_{0}|m\rangle &=& \alpha_{m}|m\rangle \nonumber\\
A^{\dagger}|m\rangle &=& N_{m}|m+1\rangle \nonumber\\
A|m\rangle &=& N_{m-1}|m-1\rangle, \label{gha1}
\end{eqnarray}
where $\alpha_{m} = g^{(m)}(\alpha_{0})$ is the $m^{th}$ iteration of $\alpha_{0}$ under $g$ and $N_{m}^{2}=\alpha_{m+1}-\alpha_{0}$. In\cite{gha1}, it was shown that eigen-energies of any quantum systems obey the equation
 \begin{eqnarray}
\tilde{\epsilon}_{n+1} &=& g(\tilde{\epsilon}_{n})
\end{eqnarray}
where $\tilde{\epsilon}_{n+1}$ and $\tilde{\epsilon}_{n}$ are the eigen-energy of successive energy levels and $g(x)$ is the characteristic function of the particular quantum system that satisfies the GHA. For example, one can obtain the standard harmonic oscillator with linear characteristic function $g(x)= x+1 $, the $q$-deformed oscillator with $g(x) = qx+1$ and free particle in an infinite square well with $g(x)=(\sqrt{x}+\sqrt{1/2})^2$. In general, GHA may not refer to smooth deformation of Heisenberg algebra\cite{gha1,gha2,gha3}.

Klauder's coherent states are by construction the eigenstates of the family of annihilation operators 
\begin{eqnarray}
A(\gamma) &=& e^{-i\gamma H/(\hbar\omega)} A\ e^{i\gamma H/(\hbar\omega)} \nonumber\\
A(\gamma)|z,\gamma\rangle &=& z |z,\gamma\rangle 
\end{eqnarray}
where $H\equiv J_0$ is the Hamiltonian of the physical system under consideration and $J\equiv|z|^2 \geq 0$ is the average energy in the elementary quantum unit of $\hbar\omega$. $z$ is the complex eigenvalue of the annihilation operators whereas $\gamma$ is the real parameter associated with the classical action angle variable\cite{gazeau2}. 

The (temporally stable) Gazeau-Klauder's generalized coherent states (GKCs) are defined as\cite{gazeau2}
\begin{eqnarray}
|J,\gamma\rangle &=& \frac{1}{N(J)}\sum_{n\geq 0} \frac{J^{(n/2)} e^{-i \gamma \epsilon_{n}}}{\sqrt{\rho_{n}}}|n\rangle\ , \ \  \rho_{n}:=\prod_{k=1}^n{\epsilon_k} \label{gkcs}
\end{eqnarray}  
where we have denoted $\epsilon_{n} = \dfrac{\tilde{\epsilon}_{n}-\tilde{\epsilon}_{0}}{\hbar\omega}$ and $|n\rangle$ is the number state. $N(J)$ is a normalization constant. For consistency, we set $\rho_{0}=1$. Note that, to ensure that both $(J,\gamma)$ are action angle variables, we need the GKCs to satisfy
\begin{eqnarray}
\langle J,\gamma|H|J,\gamma\rangle = \hbar\omega J. \label{consistentJ}   
\end{eqnarray}
We have the time independent of expectation value (temporally stable) of Hamiltonian in state with $(J,\gamma)$.

The GCs are said to be Gazeau-Klauder's type if they satisfy the following conditions:

(I) \underline{Normalizability:}
\begin{eqnarray}
\bigl|\langle J,\gamma| J,\gamma\rangle\bigr|^2 &=& 1
\end{eqnarray}

(II) \underline{Continuity in the label:}
\begin{eqnarray}
|J-J'|\Rightarrow 0\ ;\  \parallel |J,\gamma\rangle-|J',\gamma\rangle\parallel \Rightarrow 0
\end{eqnarray}

(III) \underline{Completeness:}
\begin{eqnarray}
\int (d^2 z) w(z,\gamma) |z,\gamma\rangle \langle z,\gamma|=1
\end{eqnarray}
where $(d^2 z) w(z,\gamma)$ is the measure on the Hilbert space spanned by $|z,\gamma\rangle$.

Normalization constant can be expressed as,
\begin{eqnarray}
\bigl|\langle J,\gamma| J,\gamma\rangle\bigr|^2 &=& 1\Rightarrow N(J)^2=\sum_{n\geq 0}^{\infty}\frac{J^n}{\rho_{n}}.
\end{eqnarray}
Strictly speaking, the GCS exists only if the radius of convergence 
\begin{eqnarray}
R &=& \lim_{n\rightarrow \infty}\sup\sqrt[n]{\rho_{n}}
\end{eqnarray}
is nonzero\cite{klauder1,klauder2,klauder3}. In fact, different choices of $\rho_{n}$ and hence the characteristic function $g(\tilde{\epsilon}_{n})$ give rise to many different family of GCs. On the other hand, the temporal stability condition of the eigenstates can be obtained by
\begin{eqnarray}
e^{-iHt/\hbar}|z,\gamma\rangle &=&|z,\gamma+\omega t\rangle.
\end{eqnarray}

\subsection{Gazeau-Klauder Schr$\ddot{\text{o}}$dinger Cat States (GKSCs)}

Schr$\ddot{\text{o}}$dinger Cat states (SCs) $|\psi_{sc}\rangle$ are defined as the coherent superposition of the two coherent states $|z \rangle$ and $|-z \rangle$. They have been studied well and their characteristics are summarized in the literatures\cite{klauderB,scully,mandel}. Although SCs are constructed by coherent state, they are generally nonclassical states where $|z \rangle$ and $|-z \rangle$ can be macroscopically distinguished for sufficient large $|z|$.

Schr$\ddot{\text{o}}$dinger Cat states are defined as 
\begin{eqnarray}
|\psi_{sc}\rangle = N_{sc} \bigl(|z \rangle + e^{i\phi}|-z \rangle\bigr) \label{sc}
\end{eqnarray}
where $\phi$ is the relative phase (can be taken on $[0, 2\pi]$) and $N_{sc}$ is the normalization constant. For $\phi=0$, we have the so-called even cat states which exhibit vanishing odd number probability distribution $P_n^{(\textrm{odd})}=|\langle n|\psi_{sc}\rangle|^2 = 0$. In contrast, for $\phi= \pi$, odd cat states which exhibit vanishing even number probability distribution $P_n^{(\textrm{even})} = 0$ are obtained. This is one of the most interesting statistical behaviors of SCs and we will examine the deformed-GKSCs with these statistical properties in the subsequent sections. 

With the same token, we can construct Gazeau-Klauder Schr$\ddot{\text{o}}$dinger Cat States (GKSCs).
By letting $z=\sqrt{J} e^{-i\gamma}$, we have \cite{dajka}
\begin{eqnarray}
|\psi_{gksc}\rangle = N_{gksc} \bigl(|J,\gamma \rangle + e^{i\phi}|J,\gamma +\pi \rangle\bigr) \label{gksc1}
\end{eqnarray}
where we have denoted $|z\rangle = |J,\gamma \rangle$ and $|-z\rangle = |J,\gamma+\pi\rangle$. We further substitute \eqref{gkcs} and
obtain \cite{dajka}
\begin{eqnarray}
|\psi_{gksc}\rangle = N_{gksc} \sum_{n\geq 0}^{\infty}\left\{\frac{J^{\frac{n}{2}} e^{-i\gamma \epsilon_n}}{\sqrt{\rho_n}} \Bigl[1+ e^{i(\phi-\epsilon_n \pi)}\Bigr]|n\rangle\right\} \label{gksc2}
\end{eqnarray}
where $[N_{gksc}]^{-2}= 2 \sum_{n\geq 0}^{\infty} \Bigl(\dfrac{J^n}{\rho_n}[1+ \cos(\phi-\epsilon_n \pi)]\Bigr)$.

\section{GKSCs with Minimum Length (ML) and/or Maximum Momentum (MM)}

Next, we introduce quantum deformation to Heisenberg algebra. Consider one dimension, such deformation can be generally expressed as following modified commutation relation (MCRs)
\begin{eqnarray}
[X,P]=i\hbar f(X,P) \label{mcr}
\end{eqnarray}
where $f(X,P)$ captures the position and/or momentum dependence deformation and hence play the crucial role in determine the modified dynamics. For $f(X,P) = 1$, we recover the standard quantum mechanics. Different quantum deformation models are distinguished by the expression of $f(X,P)$ and this operator-valued function is essential to study different phenomenological effects of the deformation models. There are two important phenomenological models of quantum deformation in the literature, namely the minimum length model which is inspired by string theory, black hole physics and loop quantum gravity and the maximum momentum model which is inspired by doubly special relativity. For review on generalized uncertainty principle induced by modified commutation relation, see\cite{tawfik, sabine}.

\subsection{Deformation with Minimal Length}

In literature, one of the most interesting and nontrivial model to be considered is the Kempf-Mangano-Mann (KMM) model\cite{kempf1} which satisfies the relations\footnote{KMM model can be regarded as 3-dimensional realization of Snyder model\cite{snyder} which was the first attempt to study Lorentz invariant discrete space-time back in 1940's.}
\begin{eqnarray}
[X, P] = i\hbar (1 + \tilde{\beta} P^2) ;\hspace{0.5cm} X = (1 + \tilde{\beta} p^2)x, \hspace{0.25cm} P = p \label{mcr1}
\end{eqnarray}
such that $\tilde{\beta}=\frac{\beta}{m\hbar \omega}$ with $\beta$ being the dimensionless deformed parameter and $\Bigl[\tilde{\beta}\Bigr]=$ momentum$^{-2}$. Here, both $x$ and $p$ are standard position and momentum operators which satisfy conventional commutation relation $[x, p] = i\hbar$. The modified relation exhibits an intrinsic minimum length as discussed in\cite{kempf3, bagchi} 
\begin{eqnarray}
L_{min}:= (\Delta X)\Bigr|_{min}\approx \hbar \sqrt{\tilde{\beta}} .
\end{eqnarray}

Next, we proceed to compute the deformed-GKSCs for the case of minimum length. In order to do that, we need the quantum state and eigen-energy spectrum of the deformed one dimensional simple harmonic oscillator. The perturbed Hamiltonian $H^{(ml)}$ of the deformed harmonic oscillator in noncommutative space is given by\cite{deyfringCS, deyfringR}
\begin{eqnarray} 
H^{(ml)} =\frac{P^2}{2m} + \frac{m\omega^2}{2} X^2 - \frac{\hbar \omega}{2} \left(1+\frac{\beta}{2}\right). \label{hamilSHO}
\end{eqnarray}
With representation of physical position operator $X$ given in \eqref{mcr1}, the Hamiltonian \eqref{hamilSHO} becomes non-Hermitian. Following \cite{deyfringCS,bagchi}, one can perform the Dyson map $\eta=(1 + \tilde{\beta} p^2)^{-1/2}$, whose adjoint action relates the non-Hermitian Hamiltonian in \eqref{hamilSHO} to its isospectral Hermitian
%\footnote{Alternatively, one can check the self-adjointness of the non-Hermitian operators in deformed space by deficiency indices and Von Neumann's theorem. This leads to symmetricity condition and self-adjoint extension of the physical operators. See Refs.\cite{bonneau,araujo,phdCL}} 
counterpart $h^{(ml)}$,
\begin{eqnarray} 
h^{(ml)} &=& \eta H^{(ml)}\eta^{-1}\nonumber\\
&=& \frac{p^2}{2m} + \frac{m\omega^2}{2} x^2 + \frac{\omega\beta}{4\hbar}\Bigl[p^2 x^2 + x^2 p^2 + 2x p^2 x\Bigr. \nonumber\\
&\phantom{a}& \Bigl.-2i\hbar(xp + px)\Bigr]- \frac{\hbar \omega}{2} \left(1+\frac{\beta}{2}\right) + O(\beta^2).
\end{eqnarray}
The perturbed energy spectrum is given by
\begin{eqnarray}
\frac{\epsilon^{(ml)}_n}{\hbar \omega} = n + \frac{n}{2}\bigl(n + 1\bigr)\beta + O(\beta^2). \label{mlshoE}
\end{eqnarray}
The term $O(\beta^2)$ can to be truncated from the phenomenological point of view. Notice that the deformed energy spectrum increases faster for higher $n$ when compared to the standard case. The perturbed Hamiltonian eigenstates (up to first order in $\beta$) are given by the standard nondegenerate Rayleigh-Schr$\ddot{\text{o}}$dinger perturbation theory, 
\begin{eqnarray}
|n\rangle_{ml} &=& |n^{(0)}\rangle + \sum_{k\neq n}^\infty \frac{\langle k^{(0)}|h_1|n^{(0)}\rangle}{\epsilon_n^0-\epsilon_k^0} |k^{(0)} \rangle\nonumber\\
&=& |n^{(0)}\rangle + \frac{\beta}{16} \left(\sqrt{\mathbf{P}[n+1, 4]}|n^{(0)} + 4\rangle - \sqrt{\mathbf{P}[n-3, 4]}|n^{(0)} - 4\rangle\right) \label{mlpertstate}
\end{eqnarray}
where $h_1=\omega\beta\bigl(p^2 x^2 + x^2 p^2 + 2x p^2 x - 2i\hbar(xp + px) -\hbar^2\bigr)/(4\hbar)$ is the perturbed Hamiltonian. We denoted $|n\rangle_{ml}$ and $|n^{(0)}\rangle$ as the perturbed and unperturbed Fock\rq{}s state respectively. Also, $\mathbf{P}[a, n] \equiv (a)_n$ is the Pochhammer symbol which is a compact way to express the factorial, 
\begin{eqnarray}
\mathbf{P}[a, n] \equiv (a)_n := \frac{(a + n - 1)!}{(a - 1)!}\ ; (a)_0:= 1.
\end{eqnarray}
It is clear that up to leading order in $\beta$, the perturbed eigenstates $|n\rangle_{ml}$ are not normalized in the usual manner because 
\begin{eqnarray}
_{ml}\langle m|n\rangle_{ml} = \delta_{mn} + \frac{\beta}{16}\Bigl\{(n+1)_4 \delta_{m,n+4}- (n-3)_4 \delta_{m,n-4} + \text{terms $m\leftrightarrow n$}\Bigr\}. 
\end{eqnarray}
Next, following\cite{sakuraiqm} we can renormalize the perturbed Fock\rq{}s state such that $\langle \xi^{(ml)}_m\Bigl|\xi^{(ml)}_n\rangle=\delta_{mn}$ by defining 
\begin{eqnarray}
\Bigl|\xi^{(ml)}_n\rangle:= \sqrt{Z_n^{(ml)}} |n\rangle_{ml}. \label{mlpertstate-norm}
\end{eqnarray}
The constant $Z_n^{(ml)}$ satisfies $Z_n^{(ml)}=|\langle n^{(0)}|\xi^{(ml)}_n\rangle|^2$ and thus can be regarded as the probability for the perturbed eigenstates to be found in the corresponding unperturbed eigenstate. Explicitly, it can be written as
\begin{eqnarray}
Z_n^{(ml)} \approx 1 - \sum_{k\neq n}^\infty \frac{\bigl|\langle k^{(0)}|h_1|n^{(0)}\rangle\bigr|^2}{(\epsilon_n^0-\epsilon_k^0)^2}. 
\end{eqnarray}

Following GK-approach in Sect. 2.1, the KMM-deformed GKSCs are constructed as
\begin{eqnarray}
|\psi^{(ml)}_{gksc}\rangle:= N^{(ml)}_{gksc} \sum_{n\geq 0}^{\infty}\left\{\frac{J^{\frac{n}{2}} e^{-i\gamma \epsilon^{(ml)}_n}}{\sqrt{\rho^{(ml)}_n}} \Bigl[1+ e^{i(\phi-\epsilon^{(ml)}_n \pi)}\Bigr]|\xi^{(ml)}_n\rangle\right\} \label{gkscml}
\end{eqnarray}
with
\begin{eqnarray}
\rho^{(ml)}_n := \prod_{k\geq 1}^n \epsilon^{(ml)}_k = \prod_{k\geq 1}^n \left[k+\frac{k(k+1)\beta}{2}\right] = \frac{1}{2^n}\beta^n n! \mathbf{P}[2+2/\beta, n] 
\end{eqnarray}
where $N^{(ml)}_{gksc}$ denotes the overall normalization constant and $\rho^{(ml)}_n$ is the probability distribution. Note that for undeformed case $\beta=0$, we have the Poisson distribution, $\rho_n = n!$.

\subsection{Deformation with Maximum Momentum}

Next, we consider another modified commutation relation which exhibits an intrinsic maximum momentum and particularly favored by doubly special relativity \cite{magueijo1,magueijo2,cortes,camelia} and anti-Snyder model\cite{mignemi,antisnyder}. One of such models is the Ali-Das-Vagenas (ADV) model\cite{das1, chingspectra}. In one dimension, it is described by following MCR
\begin{eqnarray}
[X,P] = i\hbar (1-\widetilde{\alpha} P)^2 ; \hspace{0.5cm} X = x, \hspace{0.25cm} P = \frac{p}{1+\widetilde{\alpha}p} \label{mcr2}
\end{eqnarray}
where $\widetilde{\alpha} = \frac{\alpha_0}{ M_{pl}c} = \frac{\alpha_0 L_{pl}}{\hbar}$, such that $M_{pl}$ is the Planck mass while $L_{pl}$ is the Plank length with $[\widetilde{\alpha}^2] = [\tilde{\beta}] =$ momentum$^{-2}$. Also, $\alpha_0$ is a dimensionless constant, typically assumed to be unity \cite{das1}. The above MCR manifestly exhibits an intrinsic maximum momentum
\begin{eqnarray}
P_{max} \approx \frac{1}{\widetilde{\alpha}}
\end{eqnarray}
and the form of \eqref{mcr2} suggests that one recovers the classicality around the maximum momentum\cite{chingspectra}.

The rescaled energy spectrum for the ADV deformed harmonic oscillator is given by
\begin{eqnarray}
\frac{\epsilon^{(mm)}_n}{\hbar \omega} &=& \frac{2n \Bigl[4+ n\alpha^2 \sqrt{2(2+\alpha^4)} -(n-1)\alpha^4\Bigr]}{\bigl(\alpha^2 + \sqrt{2(2+\alpha^4)} \bigr)\Bigl[(2n+1)\alpha^2 + \sqrt{2(2+\alpha^4)}\Bigr]^2}\nonumber\\
&=& n - \frac{3n}{2}\bigl(n + 1\bigr)\alpha^2 + O(\alpha^4) \label{mmshoE}
\end{eqnarray}
where $\alpha=\sqrt{m\hbar \omega} \widetilde{\alpha}$ is the dimensionless deformation parameter. Comparing \eqref{mmshoE} to the KMM model \eqref{mlshoE}, the ADV deformed bound state energy spectrum has a negative energy correction. This is the essential difference between these two models and it should be generic to all quantum mechanical bounded systems.

The (unnormalized) perturbed ADV-deformed Hamiltonian eigenstates (up to first-order in $\alpha$) are given by
\begin{eqnarray}
|n\rangle_{mm} &=& |n^{(0)}\rangle - \frac{i\alpha}{3\sqrt{8}} \left[\sqrt{\mathbf{P}[n+1, 3]}|n^{(0)} + 3\rangle + \sqrt{\mathbf{P}[n-2, 3]}|n^{(0)} - 3\rangle\right.\nonumber\\
&&\hspace{2cm}\left. - 9\Bigl( (n+1)^{3/2} |n^{(0)} + 1\rangle + n^{3/2} |n^{(0)} - 1\rangle\Bigr) \right] \label{mmpertstate}
\end{eqnarray}
and we can obtain the renormalized perturbed state as $\langle \xi^{(mm)}_m\Bigl|\xi^{(mm)}_n \rangle = \delta_{mn}$ through the definition $\Bigl|\xi^{(mm)}_n\rangle:= \sqrt{Z_n^{(mm)}} |n\rangle_{mm} $. It follows that the ADV-deformed GKSCs can be constructed as
\begin{eqnarray}
|\psi^{(mm)}_{gksc}\rangle:= N^{(mm)}_{gksc} \sum_{n\geq 0}^{\infty}\left\{\frac{J^{\frac{n}{2}} e^{-i\gamma \epsilon^{(mm)}_n}}{\sqrt{\rho^{(mm)}_n}} \Bigl[1+ e^{i(\phi-\epsilon^{(mm)}_n \pi)}\Bigr]|\xi^{(mm)}_n\rangle\right\} \label{gkscmm}
\end{eqnarray}
with
\begin{eqnarray}
\rho^{(mm)}_n := \prod_{k\geq 1}^n \epsilon^{(mm)}_k = \frac{n! (2\mu -1)^n \mathbf{P}[2(\mu +1), n]}{(2\alpha^2)^n (2\mu +1)^n \bigl(\mathbf{P}[\mu + \frac{3}{2}, n]\bigr)^2} 
\end{eqnarray}
where we denote $\mu=\sqrt{\frac{1}{2}+\frac{1}{\alpha^4}}$ and $N^{(mm)}_{gksc}$ is the overall normalization constant. Also, note that for $\rho^{(mm)}_n\bigr|_{\alpha\rightarrow 0} = n!$ we recover the Poisson distribution.

\section{Probability Distribution and Entropy of the Deformed-GKSCs}

\subsection{Normalization Constant}

To ensure the orthonormality condition $\bigl|\langle \psi^{(ml)}_{gksc}|\psi^{(ml)}_{gksc}\rangle\bigr|^2 = 1 = \bigl|\langle \psi^{(mm)}_{gksc}|\psi^{(mm)}_{gksc}\rangle\bigr|^2$ of the deformed-GKSCs \eqref{gkscml} and \eqref{gkscmm} the normalization constant $N(J)$ is fixed as the follows. The normalization constant for KMM model is given by
\begin{eqnarray}
N^{(ml)}_{gksc} (J,\beta,\phi) = \left(2\sum^{\infty}_{n=0} \frac{2^n J^n}{\beta^n n! \mathbf{P}[2+2/\beta, n]} \Bigl[1+\cos (\phi - \epsilon^{(ml)}_n \pi)\Bigr]\right)^{-1/2}
\end{eqnarray}
and for ADV model is 
\begin{eqnarray}
N^{(mm)}_{gksc} (J,\alpha,\phi) = \left(2\sum^{\infty}_{n=0} \frac{2^n J^n \mathbf{P}[\mu+3/2, n]}{n!\alpha^{2n} \mathbf{P}[2(\mu+1), n]} \Bigl[1+\cos (\phi - \epsilon^{(mm)}_n \pi)\Bigr]\right)^{-1/2} .
\end{eqnarray}
 
\begin{figure}[h!]
    \centering
    \begin{subfigure}[b]{0.485\textwidth}
        \includegraphics[width=\textwidth]{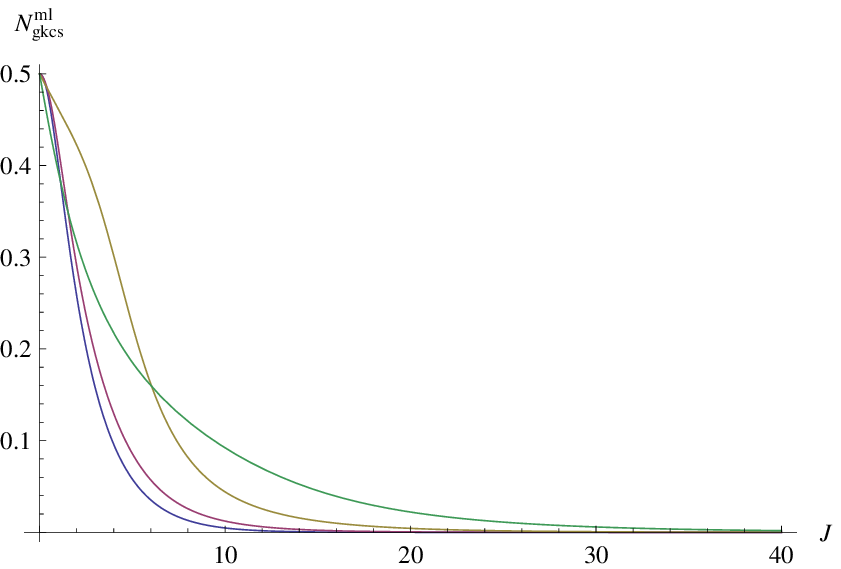}
        \caption{$N^{(ml)}_{gksc}$ for the minimum length (KMM) model corresponding to different values of deformed parameter $\beta$. Purple, gold and green lines are the deformed cases with $\beta = (0.1, 0.3, 0.7)$ respectively.}
        \label{fig2a}
    \end{subfigure}
    ~ %add desired spacing between images, e. g. ~, \quad, \qquad, \hfill etc. 
    \begin{subfigure}[b]{0.485\textwidth}
        \includegraphics[width=\textwidth]{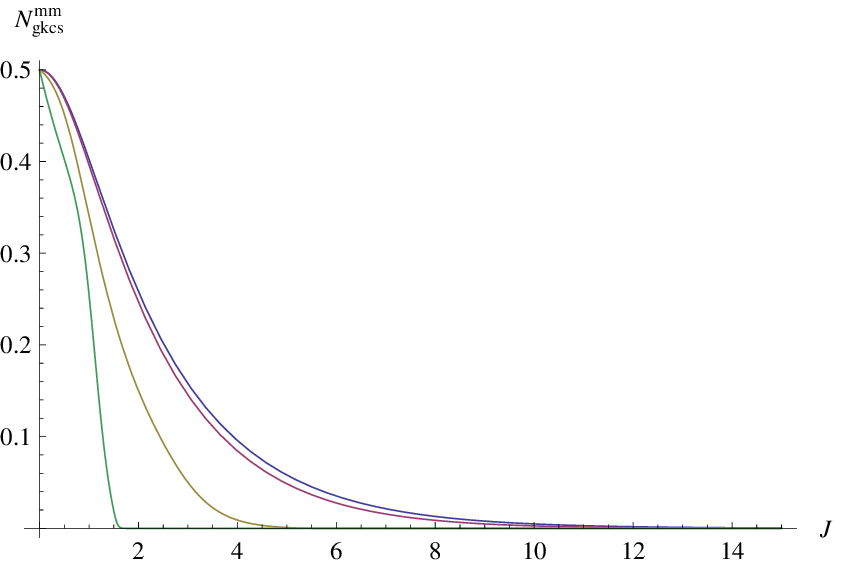}
        \caption{$N^{(mm)}_{gksc}$ for the maximum momentum (ADV) model corresponding to different values of deformed parameter $\alpha$. Purple, gold and green lines are the deformed cases with $\alpha = (0.1, 0.3, 0.5)$ respectively.}
        \label{fig2b}
    \end{subfigure}
    \caption{The normalization constant $N_{gksc}$ for both deformation case with $\phi = 0$ (even cat states). Blue line corresponds to undeformed case.}\label{fig: Normalize_even}
\end{figure}

In Fig.(1), we plot both $N^{(ml)}_{gksc}$ and $N^{(mm)}_{gksc}$ for even GKSCs ($\phi = 0$) as the function of average energy $J$ for different values of deformation parameter $\beta$ and $\alpha$. As an approximation, we truncate the summation at $n=100$ for KMM models and $n=120$ for ADV model. In both cases, we recover the undeformed normalization constant $N_{gksc}(J,\phi)=\bigl(2(e^J + e^{-J}\cos\phi)\bigr)^{-1/2}$ as $(\beta,\alpha)\rightarrow 0$. In the KMM model, it is observed that $N^{(ml)}_{gksc}$ tends to increase with increasing $\beta$. In contrast, $N^{(mm)}_{gksc}$ in ADV model tends to decrease with increasing $\alpha$. Note that from phenomenological point of view, the physically acceptable range of deformed parameter should be very small, $|\beta|, |\alpha| <<1$. 

\subsection{Probability Distribution}

It is well known that Schr$\ddot{\text{o}}$dinger Cat states (SCs) have very specific statistical characteristics. In standard quantum optics, the even SCs ($\phi=0$) have vanishing probability of detecting an odd number of photons while the odd SCs ($\phi = \pi$) have vanishing probability of detecting an even number of photons. It is interesting to find out whether the deformed GKSCs still possess these important statistical features. The probability distribution is defined as
\begin{eqnarray}
P_n(J,\phi)&:=& |\langle n|\psi_{sc}\rangle|^2.
\end{eqnarray}
For the KMM model, the probability distribution is given by
\begin{eqnarray}
P^{(ml)}_n(J,\phi,\beta) &=& \left|\langle \xi_{n}^{(ml)}|\psi_{gksc}^{(ml)}\rangle\right|^2\nonumber\\
&=& 2\bigl(N^{(ml)}_{gksc}\bigr)^2 \frac{(2J)^n}{\beta^n n!\ \mathbf{P}[2+2/\beta, n]} \Bigl[1+\cos (\phi - \epsilon^{(ml)}_n \pi)\Bigr] \label{ProbDKMM}
\end{eqnarray}
and we plot the probability distribution function as function of $J$ and $n$ respectively. For the even (odd) states, we set $\phi=0$ ($\phi=\pi$). 
\begin{figure}[h!]
    \centering
    \begin{subfigure}[b]{0.48\textwidth}
        \includegraphics[width=\textwidth]{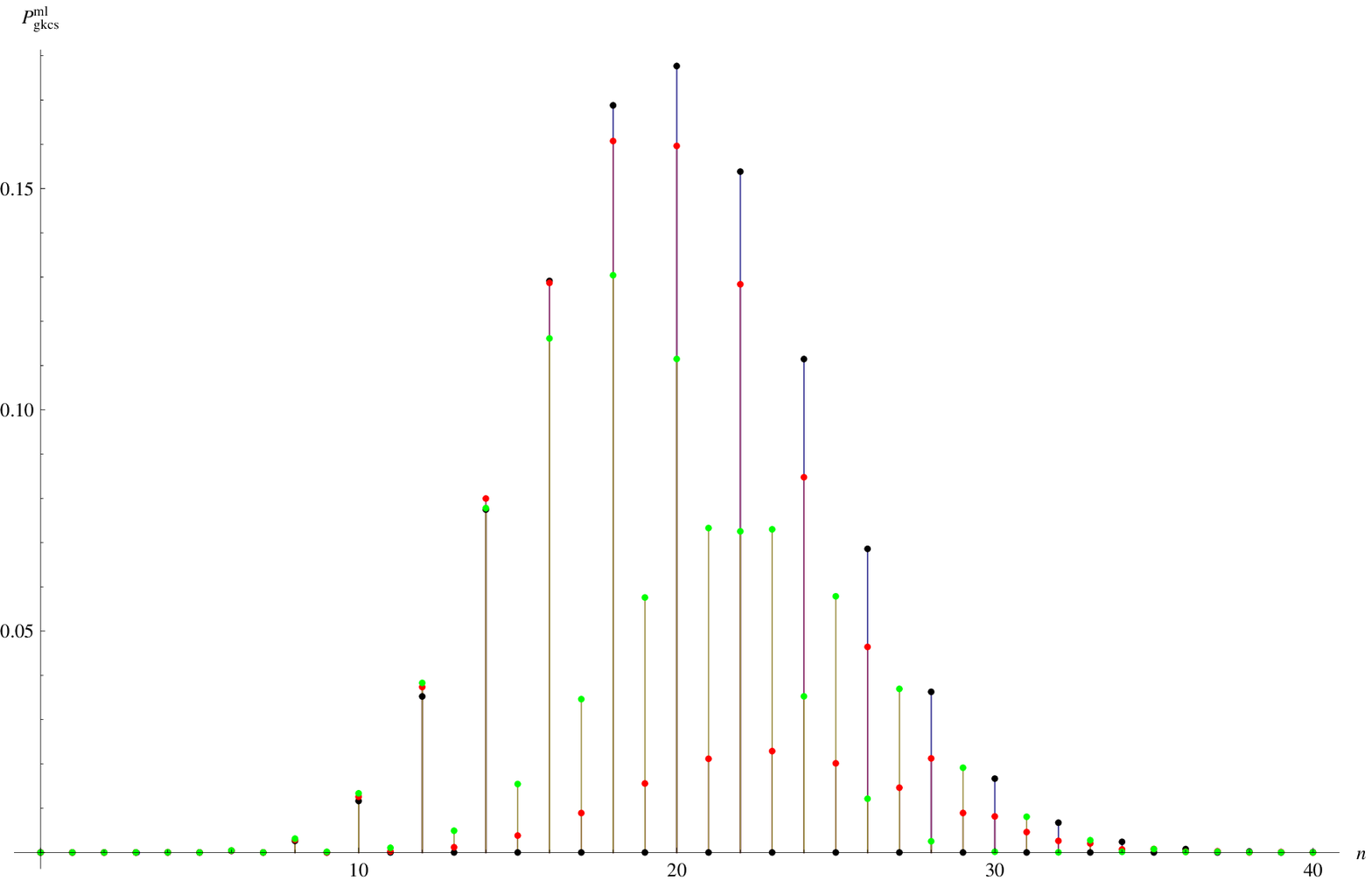}
        \caption{$P^{(ml)}_n (J, 0, \beta)$ as function of $n$ for $J=20$. Black dot is the undeformed case while red and green dots are the deformed cases with $\beta = (0.001, 0.002)$ respectively.}
        \label{fig2a}
    \end{subfigure}
    ~ %add desired spacing between images, e. g. ~, \quad, \qquad, \hfill etc. 
    \begin{subfigure}[b]{0.48\textwidth}
        \includegraphics[width=\textwidth]{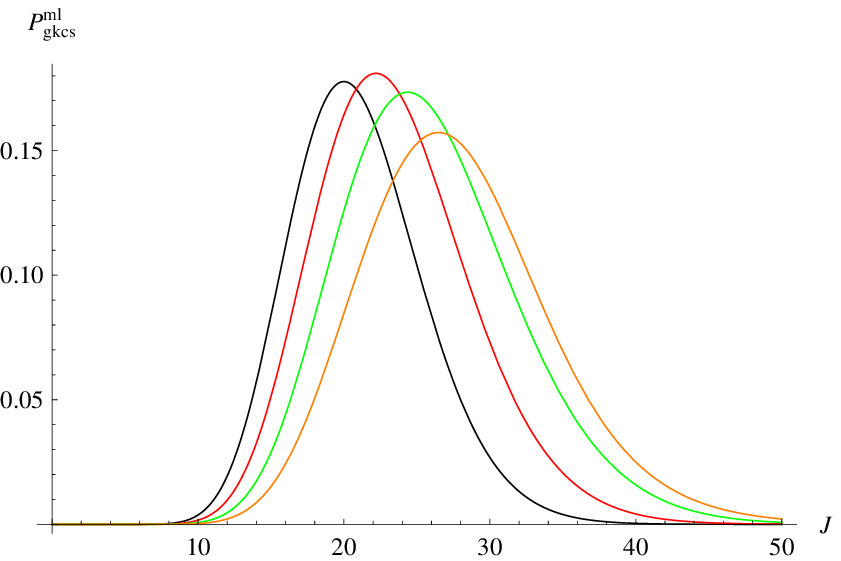}
        \caption{$P^{(ml)}_n (J, 0, \beta)$ as function of $J$ for $n=20$. Black line is the undeformed case while red, green and orange lines are the deformed cases with $\beta = (0.001, 0.002, 0.003)$ respectively.}
        \label{fig2b}
    \end{subfigure}
    \caption{The probability distribution $P^{(ml)}_n (J, \phi, \beta)$ for the minimum length (KMM) model with $\phi = 0$ corresponding to different values of deformation parameter $\beta$.}\label{fig: PKMM_even}
\end{figure}

\begin{figure}[h!]
    \centering
    \begin{subfigure}[b]{0.48\textwidth}
        \includegraphics[width=\textwidth]{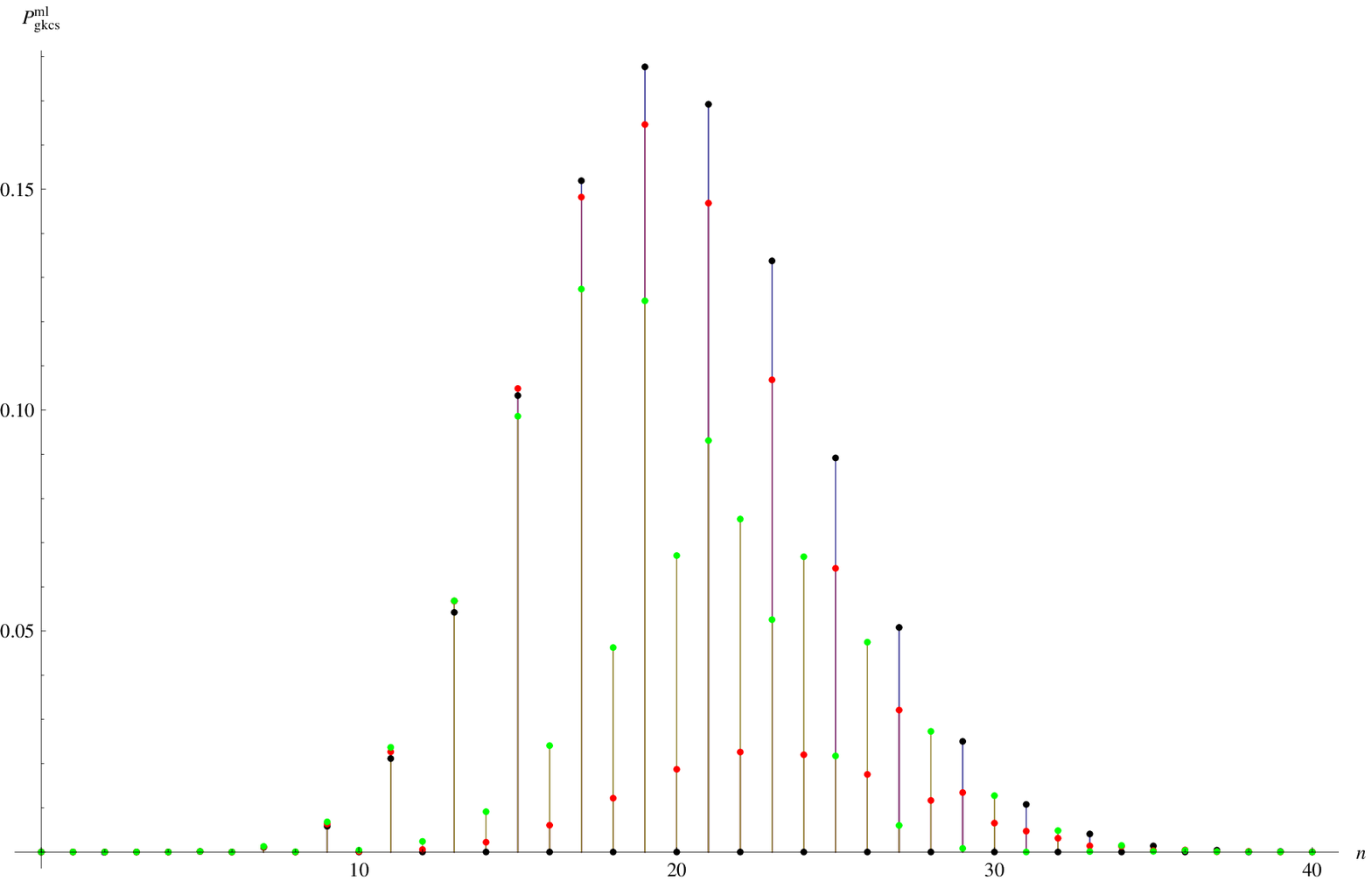}
        \caption{$P^{(ml)}_n (J, \pi, \beta)$ as function of $n$ for $J=20$. Black dot is the undeformed case while red and green dots are the deformed cases with $\beta = (0.001, 0.002)$ respectively.}
        \label{fig3a}
    \end{subfigure}
    ~ %add desired spacing between images, e. g. ~, \quad, \qquad, \hfill etc. 
    \begin{subfigure}[b]{0.48\textwidth}
        \includegraphics[width=\textwidth]{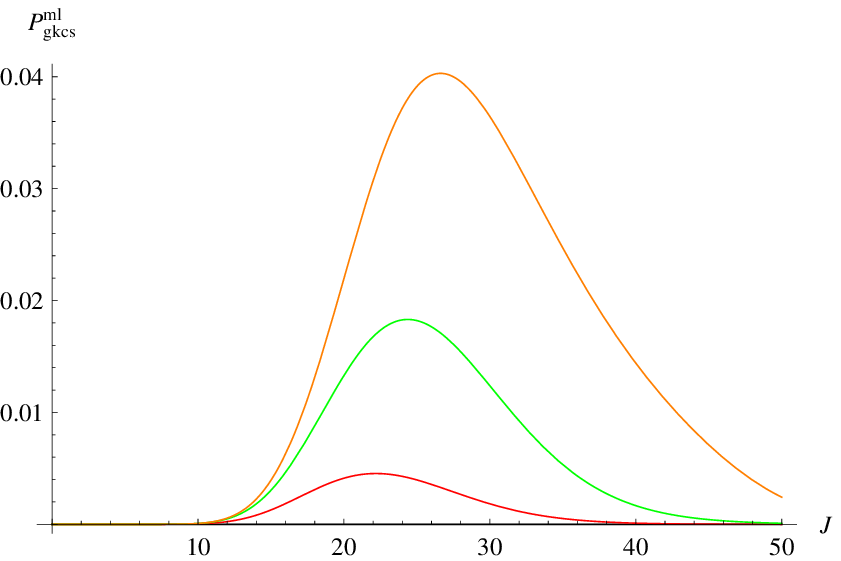}
        \caption{$P^{(ml)}_n (J, \pi, \beta)$ as function of $J$ for $n=20$. Black line is the undeformed case while red, green and orange lines are the deformed cases with $\beta = (0.001, 0.002, 0.003)$ respectively.}
        \label{fig3b}
    \end{subfigure}
    \caption{The probability distribution $P^{(ml)}_n (J, \phi, \beta)$ for the minimum length (KMM) model with $\phi = \pi$ corresponding to different values of deformation parameter $\beta$.}\label{fig: PKMM_odd}
\end{figure}

From Fig.(2a) and Fig.(3a), we observed that KMM-deformed GKSCs do not possess the statistical characteristics of even (odd) SCs anymore. In Fig.(2a), by setting $J = 20$, we notice that the “even” deformed-GKSCs now possess nonvanishing odd probability distribution. Similarly, in Fig.(3a), the “odd” deformed-GKSCs possess nonvanishing even probability distribution. Consider the “even” deformed-GKSCs, when the deformation parameter $\beta$ increases, the probability of observing even $n$ numbers of photons decreases and the peak is shifted towards smaller effective $n$. In contrast, the probability of observing odd $n$ numbers of photon which is supposedly zero in standard SCs increases due to the deformation. Similar conclusion can be drawn on the “odd” deformed GKSC in Fig.(3a). As a result, for a nonzero deformation $\beta$, it induces a mixture of two probability distributions to detect odd and even GKSCs respectively. Effectively, this behavior shows up in a Kerr-type oscillator in nonlinear medium\cite{dajka}.

Next, we set the photon number $n = 20$ (even) and illustrate the probability distribution of “even/odd” KMM-deformed-GKSCs in Fig.(2b) and Fig.(3b) respectively corresponding to different values of $\beta$. It is observed that in Fig.(2b) and Fig.(3b), as the deformation increases, the probability decreases (increases) and spreads with the peak moving towards larger $J$. This can be understood from \eqref{mlshoE}, such that the deformed energy spectrum increases with increasing deformation. We expect the average energy to increase and thus the peak of probability distribution shifts towards larger $n$.

For ADV model with maximum momentum, we compute the probability distribution to be
\begin{eqnarray}
&\phantom{a}&P^{(mm)}_n(J,\phi,\alpha) = \left|\langle \xi_n^{(mm)}|\psi_{gksc}^{(mm)}\rangle\right|^2\nonumber\\
&&= 2\bigl(N^{(mm)}_{gksc}\bigr)^2 \frac{\bigl[2J\alpha^2(2\mu+1)\bigr]^n (\mathbf{P}[\mu+3/2, n])^2}{n!(2\mu-1)^n \mathbf{P}[2(\mu+1), n]}\Bigl[1+\cos (\phi - \epsilon^{(mm)}_n \pi)\Bigr]. \label{ProbDADV}
\end{eqnarray}

Similar to KMM model, we plot the probability distribution function as function of $n$ and $J$ in the following graphs for even ($\phi=0$) and odd ($\phi=\pi$) states separately.
\begin{figure}[h!]
    \centering
    \begin{subfigure}[b]{0.48\textwidth}
        \includegraphics[width=\textwidth]{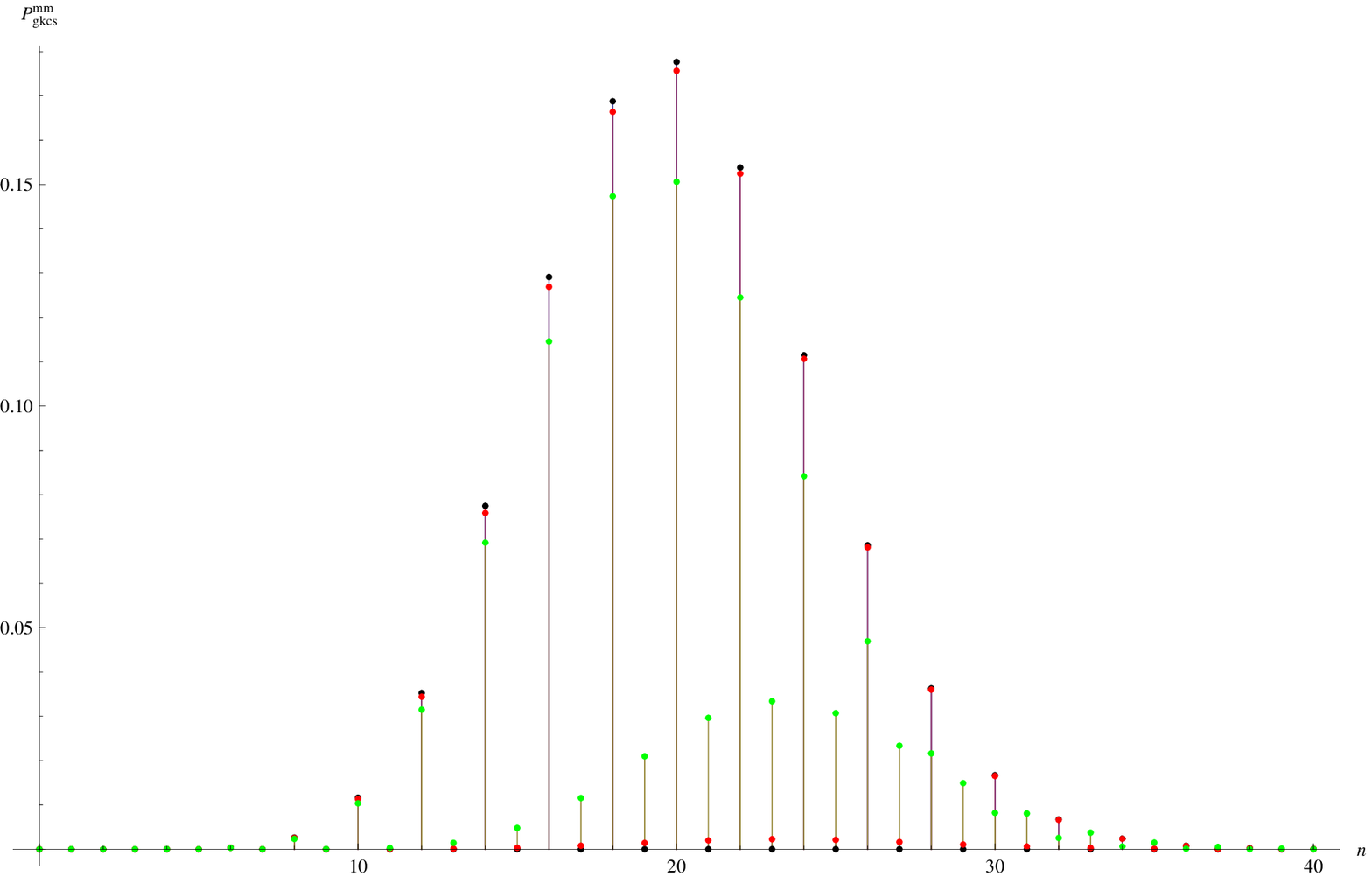}
        \caption{$P^{(mm)}_n (J, 0, \alpha)$ as function of $n$ for $J=20$. Black dot is the undeformed case while red and green dots are the deformed cases with $\alpha = (0.01, 0.02)$ respectively.}
        \label{fig4a}
    \end{subfigure}
    ~ %add desired spacing between images, e. g. ~, \quad, \qquad, \hfill etc. 
    \begin{subfigure}[b]{0.48\textwidth}
        \includegraphics[width=\textwidth]{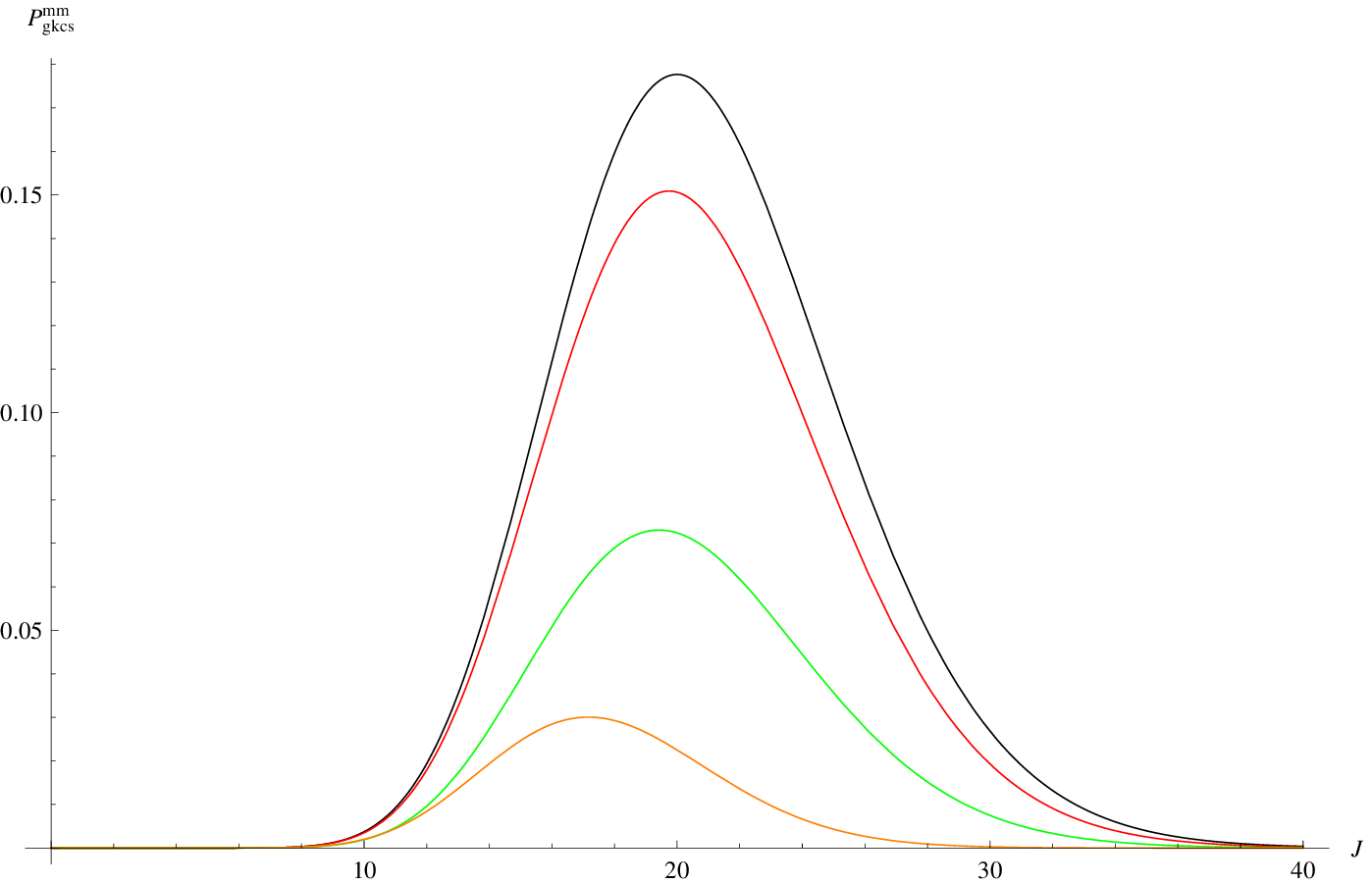}
        \caption{$P^{(mm)}_n (J, 0, \alpha)$ as function of $J$ for $n=20$. Black line is the undeformed case while red, green and orange lines are the deformed cases with $\alpha = (0.02, 0.03, 0.07)$ respectively.}
        \label{fig4b}
    \end{subfigure}
    \caption{The probability distribution $P^{(mm)}_n (J, \phi, \alpha)$ for the intrinsic maximum momentum (ADV) model with $\phi = 0$ corresponding to different values of deformed parameter $\alpha$. }\label{fig: PADV_even}
\end{figure}

\begin{figure}[h!]
    \centering
    \begin{subfigure}[b]{0.48\textwidth}
        \includegraphics[width=\textwidth]{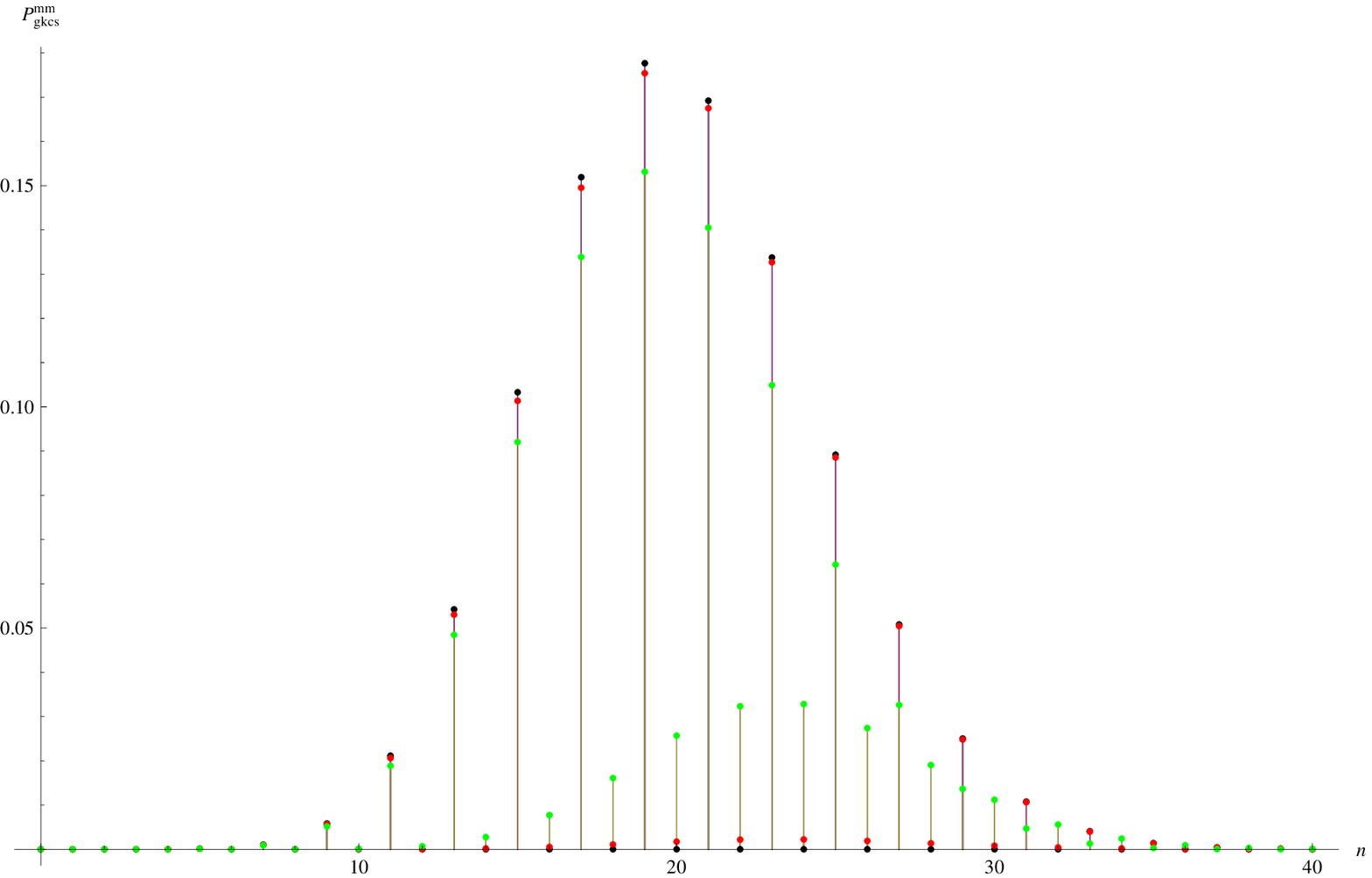}
        \caption{$P^{(mm)}_n (J, \pi, \alpha)$ as function of $n$ for $J=20$. Black dot is the undeformed case while red and green dots are the deformed cases with $\alpha = (0.01, 0.02)$ respectively.}
        \label{fig5a}
    \end{subfigure}
    ~ %add desired spacing between images, e. g. ~, \quad, \qquad, \hfill etc. 
    \begin{subfigure}[b]{0.48\textwidth}
        \includegraphics[width=\textwidth]{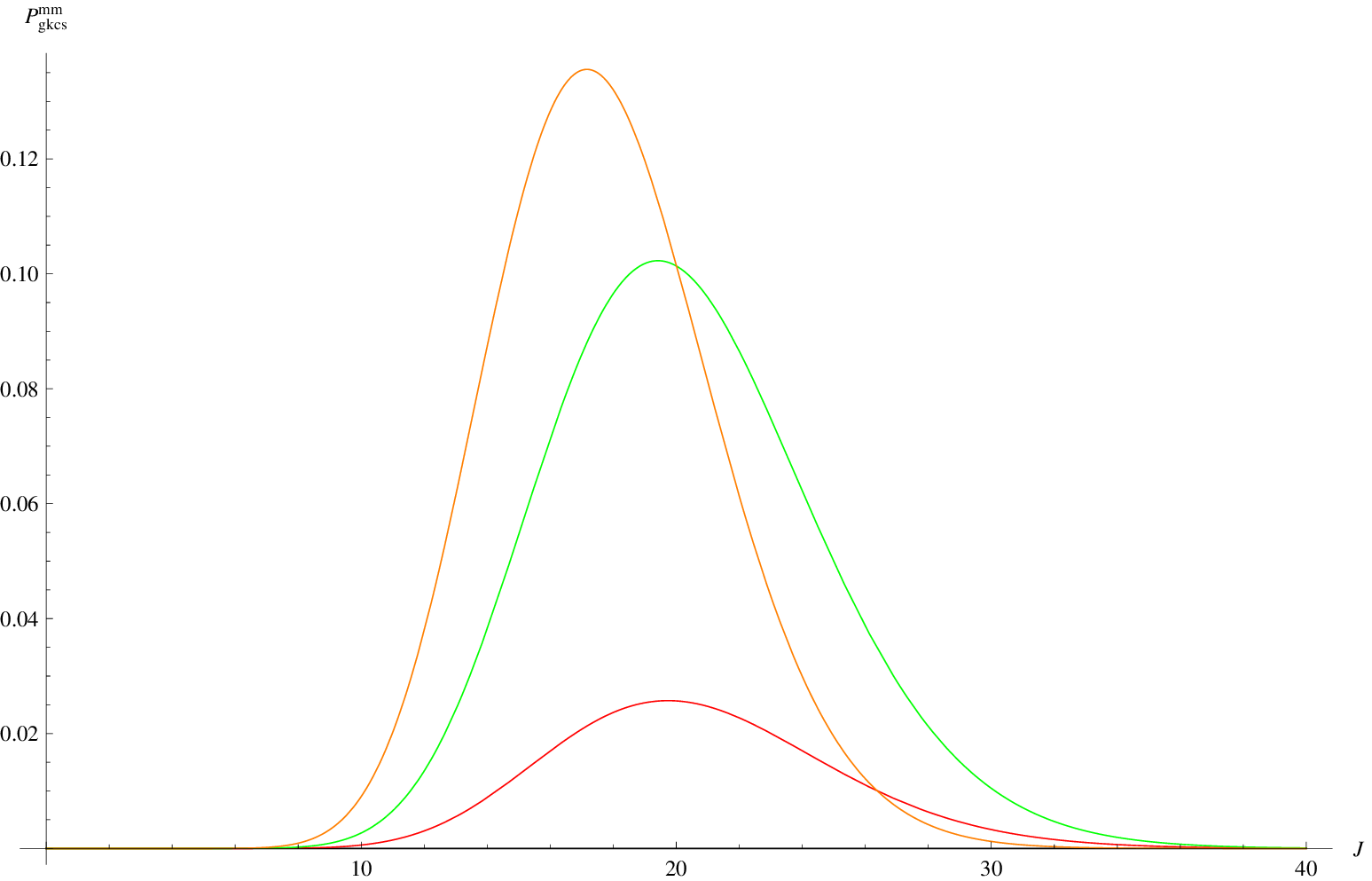}
        \caption{$P^{(mm)}_n (J, \pi, \alpha)$ as function of $J$ for $n=20$. Black line is the undeformed case while red, green and orange lines are the deformed cases with $\alpha = (0.02, 0.03, 0.07)$ respectively.}
        \label{fig5b}
    \end{subfigure}
    \caption{The probability distribution $P^{(mm)}_n (J, \phi, \alpha)$ for the intrinsic maximum momentum (ADV) case with $\phi = \pi$ corresponding to different values of deformed parameter $\alpha$.}\label{fig: PADV_odd}
\end{figure}

From Fig.(4a) and Fig.(5a), we also observe that the ADV-deformed GKSCs do not possess even and odd statistical characteristics. Similar to KMM model, the “even” ADV-deformed GKSCs now possess relatively small but nonvanishing odd probability distribution while the “odd” ADV-deformed GKSCs possess small, nonvanishing even probability distribution. By setting the photon number to be $n = 20$, we illustrate the probability distribution of the “even” and “odd” ADV-deformed GKSCs as a function of energy $J$ in Fig.(4b) and Fig.(5b) respectively. It is observed that in Fig.(4b) as the deformation increases, the probability decreases and spreads with the peak moving towards smaller $J$. Fig.(5b), for the odd states shows a contrasting trends. From \eqref{mmshoE}, we see that the deformed energy spectrum decreases with increasing deformation. We expect the average energy to decrease and thus the peak of probability distribution shifts towards smaller $n$. This is the crucial difference between the KMM model and the ADV model as pointed out in\cite{chingspectra}.

In conclusion, we see that both KMM and ADV-deformed GKSCs generally do not possess the specific statistical features of standard SCs. However, some new statistical features may appear in the deformed GKSCs. Although in general the probability distribution is nonvanishing for deformed GKSCs, we see that from \eqref{ProbDKMM}, \eqref{ProbDADV} it is possible for the term $\bigl[1+\cos (\phi - \epsilon_n \pi)\bigr]$ to vanish for certain state $n$ and certain value of deformation parameter if $\phi-\epsilon_n \pi = (2\kappa + 1) \pi$ where $k\in Z$. This new feature can be useful in the study of the deviation in photon's full counting statistics and Fano’s noise factor that is induced by different classes of modified commutation relation MCR. In principle, these two important classes of MCR can be distinguished by quantum optical experiment in laboratory setting.

\subsection{Entropy of KMM/ADV-deformed GKSCs}
Following \cite{pathria}, we define the Gibb’s entropy of the system (canonical ensemble) as the standard logarithmic measure of the density of states in the phase space given by
\begin{eqnarray}
S(J,\phi):=-k_B \sum_{n=0}^{\infty} P_n(J,\phi) \ln P_n(J,\phi) 
\end{eqnarray}
where $k_B$ is the Boltzmann’s constant. This allows us to define the entropy of KMM/ADV-deformed GKSCs in a similar way. 

\begin{figure}[h!]
    \centering
    \begin{subfigure}[b]{0.48\textwidth}
        \includegraphics[width=\textwidth]{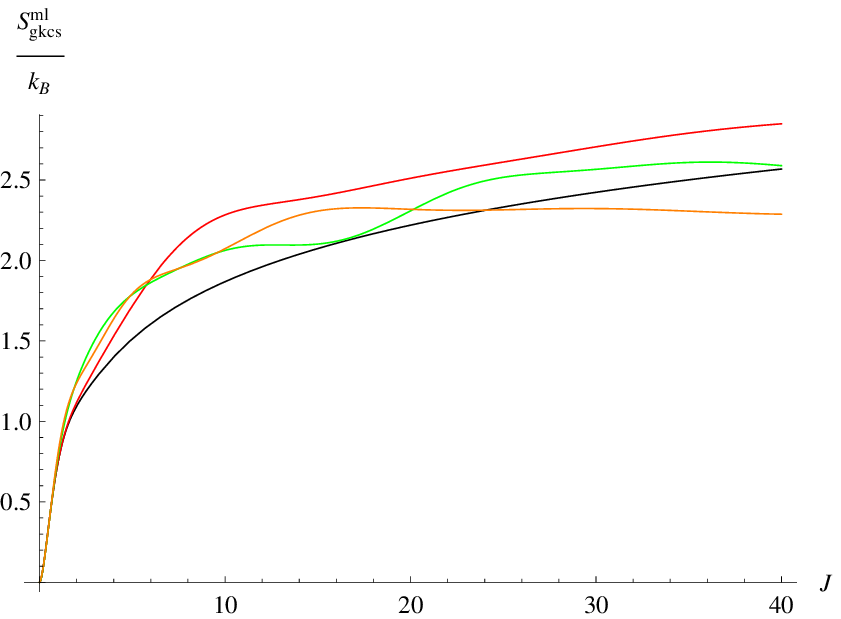}
        \caption{The entropy $S^{(ml)}$ for the KMM model with $\phi= 0$ corresponding to different values of deformed parameter $\beta$. Black line is the undeformed case while red, green and orange lines are the deformed cases with $\beta = (0.01, 0.05, 0.1)$ respectively.}
        \label{fig6a}
    \end{subfigure}
    ~ %add desired spacing between images, e. g. ~, \quad, \qquad, \hfill etc. 
    \begin{subfigure}[b]{0.48\textwidth}
        \includegraphics[width=\textwidth]{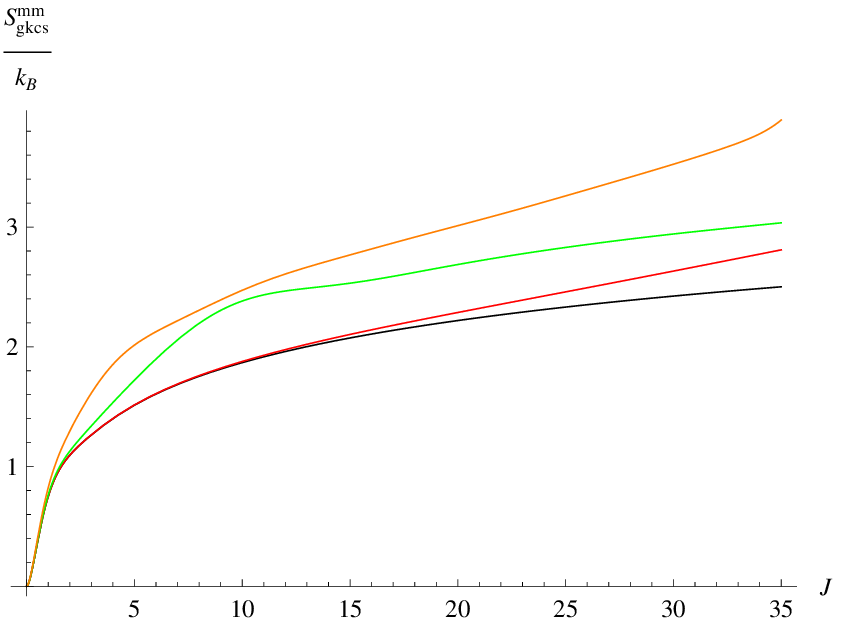}
        \caption{The entropy $S^{(mm)}$ for the ADV model with $\phi = 0$ corresponding to different values of deformed parameter $\alpha$. Black line is the undeformed case while red, green and orange lines are the deformed cases with $\alpha = (0.01, 0.05, 0.1)$ respectively.}
        \label{fig6b}
    \end{subfigure}
    \caption{Gibb's entropy for even GKSCs.}\label{fig: Entropy}
\end{figure}

We plot the entropy of both KMM model $S^{(ml)}$ and ADV model $S^{(mm)}$ for even-GKSCs as the function of the average energy $J$ in Fig.(6a) and Fig.(6b) respectively. We observe that in both models with small deformation parameters (phenomenologically preferred), the entropy of deformed GKSCs is generally increased when compared to the undeformed SCs. The difference in entropy between the two deformations is not significant for small deformation parameters and short range of average energy $J$. Similar results can be obtained for the odd-GKSCs case.

\section{Nonclassical Properties of the Deformed GKSCs}

\subsection{Photon Statistics and Number Squeezing of GKSCs}

In quantum optics, the measure of deviation from the standard Poissonian distribution is given by the famous Mandel parameter defined by
\begin{eqnarray}
Q &:=& \frac{\langle J,\gamma|(\Delta N)^2|J,\gamma\rangle}{\langle J,\gamma|N|J,\gamma\rangle} - 1 = \frac{\sum_{n=0}^\infty n^2 P_n -\bigl(\sum_{n=0}^\infty n P_n\bigr)^2}{\sum_{n=0}^\infty n P_n} -1.
\end{eqnarray}
The value of $Q$ number defines the characteristics of the quantum statistical distribution. $Q = 0$ corresponds to the standard Poissonian (classical) distribution, e.g. coherent light. For $Q > 0$, it refers to the super-Poissonian which corresponds to photon bunching statistics, e.g. thermal light. For $Q <0$, it refers to the sub-Poissonian which corresponds to photon anti-bunching statistics, e.g. squeezed coherent light. For standard SCs, it is a well known fact that even (odd) cat states exhibit photon number bunching (anti-bunching) respectively. Subsequently we would like to examine the deviation induced by the quantum deformation on this unique characteristic of the cat states. For both of our deformed KMM and ADV models, we define
\begin{eqnarray}
Q_q (J,\phi,q) &=&\frac{\sum_{n=0}^\infty n^2 P^q_n -\bigl(\sum_{n=0}^\infty n P^q_n\bigr)^2}{\sum_{n=0}^\infty n P^q_n} -1
\end{eqnarray}
where $q = (\beta, \alpha)$ refers to the classes of deformation, either KMM or ADV models whereby the probability distributions are described by \eqref{ProbDKMM} and \eqref{ProbDADV}. 

To obtain numerical plots, we perform suitable truncation up to $n = 400$ in the calculation of $Q_{\beta}$ and $Q_{\alpha}$ respectively.
\begin{figure}[h!]
    \centering
    \begin{subfigure}[b]{0.48\textwidth}
        \includegraphics[width=\textwidth]{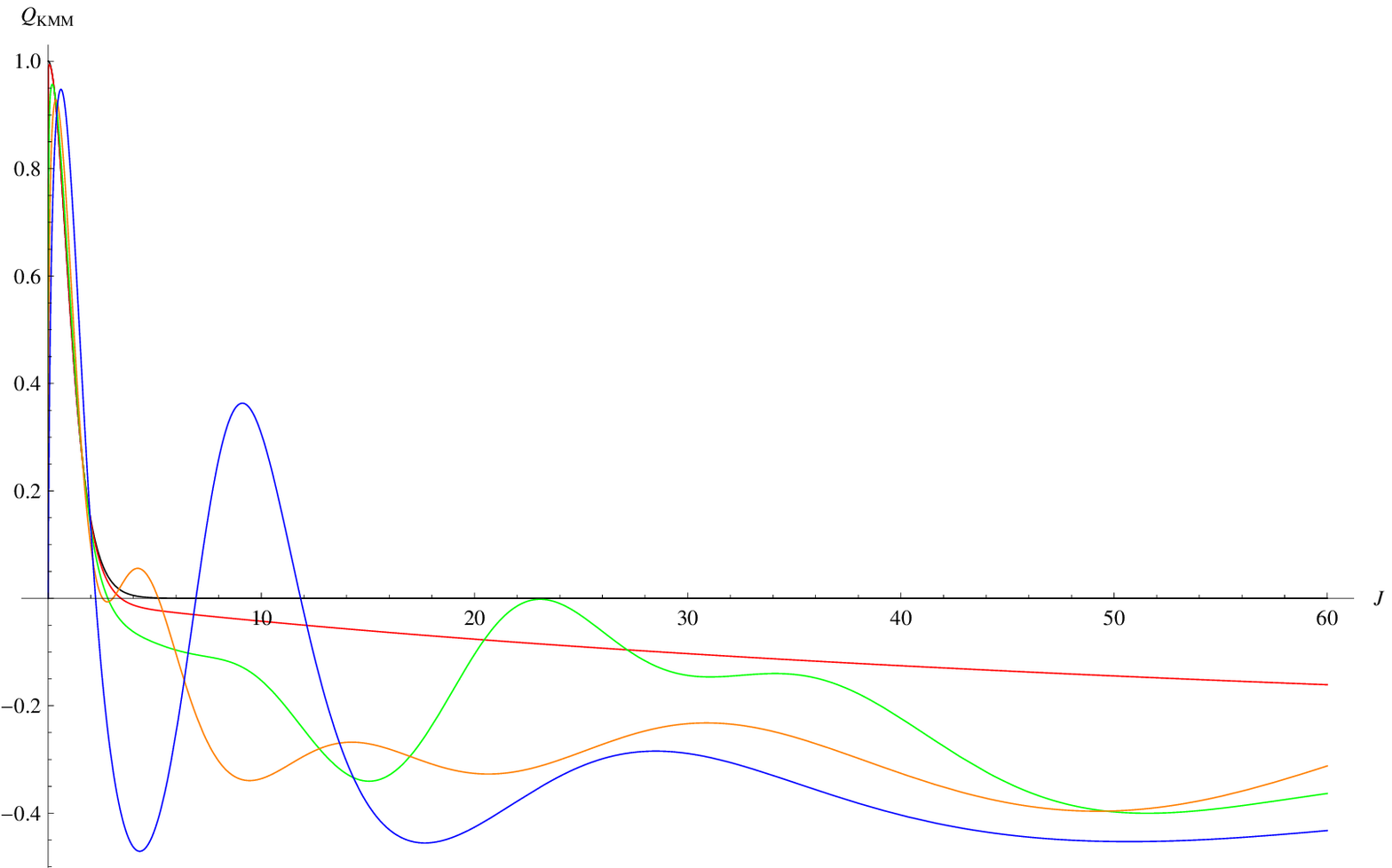}
        \caption{The Mandel parameter $Q_{\beta}$ for the even GKSCs.}
        \label{fig7a}
    \end{subfigure}
    ~ %add desired spacing between images, e. g. ~, \quad, \qquad, \hfill etc. 
    \begin{subfigure}[b]{0.48\textwidth}
        \includegraphics[width=\textwidth]{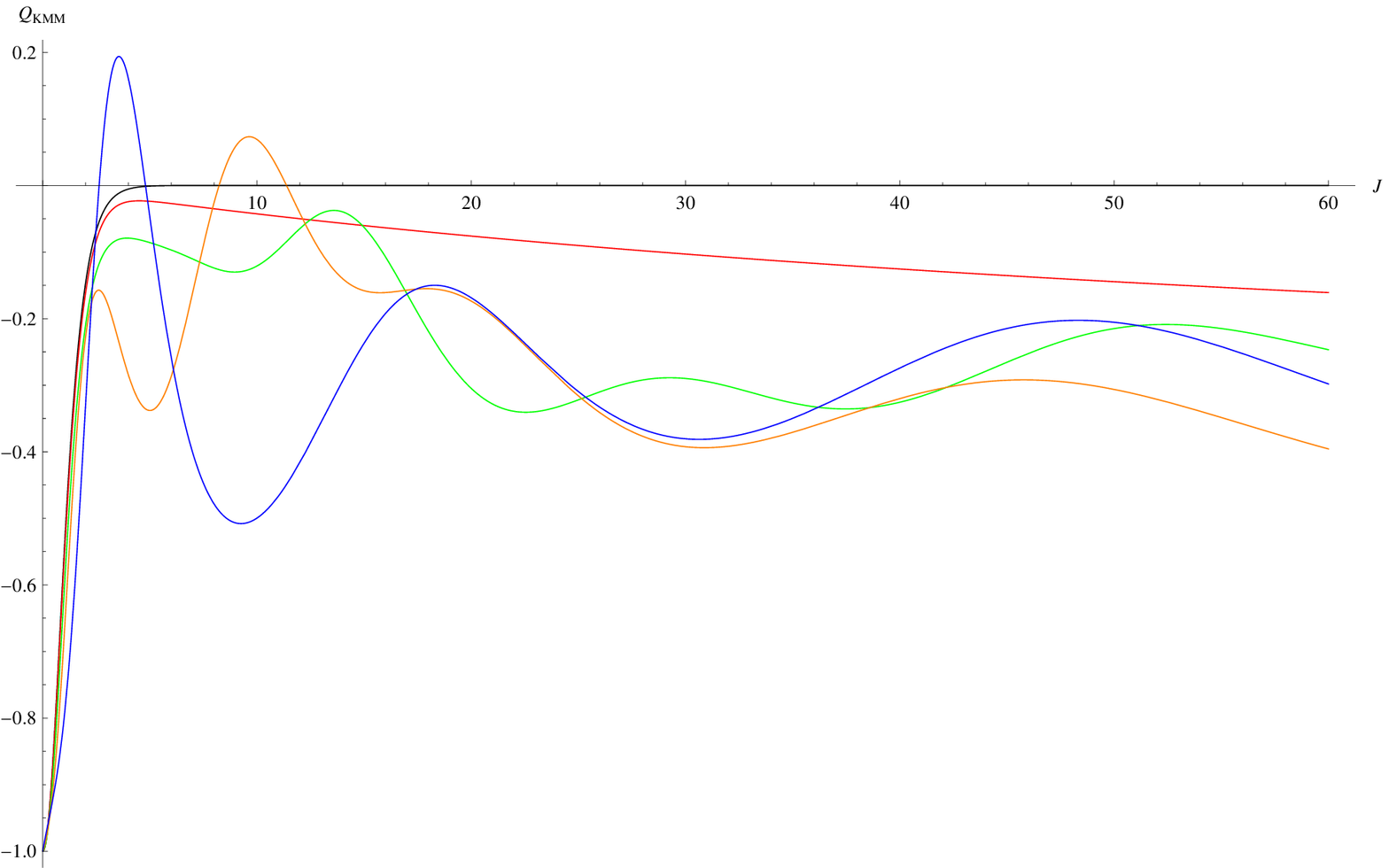}
        \caption{The Mandel parameter $Q_{\beta}$ for the odd GKSCs.}
        \label{fig7b}
    \end{subfigure}
    \caption{Mandel parameter for KMM models $Q_{\beta}$ with different values of deformation parameter $\beta$. Black line is the undeformed case while red, green, orange and blue lines are the deformed cases with $\beta = (0.01, 0.05, 0.1, 0.15)$ respectively. }\label{fig: MandelQKMM}
\end{figure}

In Fig.(7a) and Fig.(7b), we plot the Mandel's parameter $Q_\beta$ as a function of $J$ for KMM even (odd) states respectively. The illustrations show that the photon statistics of even (odd) KMM-GKSCs can be super/sub-Poissonian (positive/negative $Q$ value) for small $J$ that is similar to the undeformed standard SCs. However, both even and odd cat states tend to become more “nonclassical” sub-Poissonian (negative $Q$ value) as $J$ increases. Note that the different amount of deformation $\beta$ will only affect the oscillatory behavior of $Q_\beta$ but the $Q$-values is always bounded to be negative for increasing $J$. Thus, in KMM model, all the GKSCs get more “nonclassical” in the higher energy sector. Interestingly, the photon number distribution of the deformed even cat states experiencing more squeezing compared to standard SCs, in which the latter is always photon-bunched $Q > 0$. However, for certain values of $\beta$ and relatively small $J$, super-Poissionian is still possible.

In Fig. (8a) and Fig.(8b), we plot the Mandel's parameter $Q_\alpha$ as a function of $J$ for ADV-GKSCs even (odd) states respectively. 
\begin{figure}[h!]
    \centering
    \begin{subfigure}[b]{0.48\textwidth}
        \includegraphics[width=\textwidth]{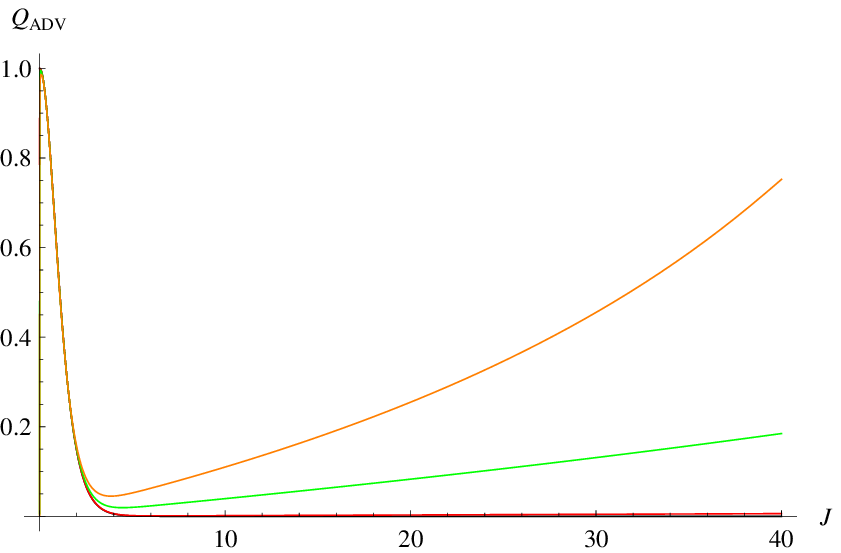}
        \caption{The Mandel parameter $Q_{\alpha}$ for the even GKSCs.}
        \label{fig8a}
    \end{subfigure}
    ~ %add desired spacing between images, e. g. ~, \quad, \qquad, \hfill etc. 
    \begin{subfigure}[b]{0.48\textwidth}
        \includegraphics[width=\textwidth]{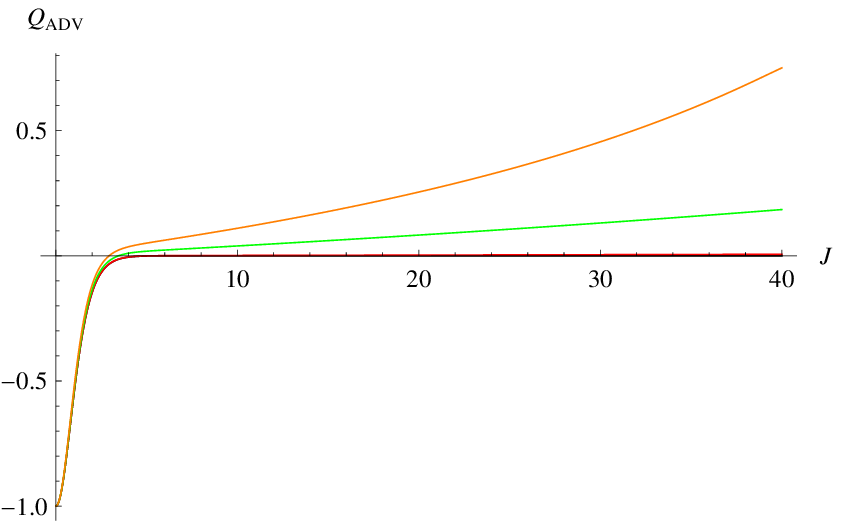}
        \caption{The Mandel parameter $Q_{\alpha}$ for the odd GKSCs.}
        \label{fig8b}
    \end{subfigure}
    \caption{Mandel parameter for ADV models $Q_{\alpha}$ with different values of deformation parameter $\alpha$. Black line is the undeformed case while red, green, and orange lines are the deformed cases with $\alpha = (0.01, 0.05, 0.08)$ respectively. }\label{fig: MandelQADV}
\end{figure}
For small $J$, both even (odd) states remained as photon bunched (anti-bunched) that is similar to the undeformed case. However, as $J$ increases, the $Q_\alpha$-values of even states still remain positive and increases with $J$ and $\alpha$ while the odd states become super-Poissionian. Thus, all of the GKSCs get more “classical” for ADV model in the higher energy sector. Note that from the phenomenological point of view, larger values of $J$ should be physically more relevant. The existence of such classicality is intimately related to the vanishing MCR in \eqref{mcr2} as $P\rightarrow P_{max}=\frac{1}{\widetilde{\alpha}}$.

In comparison between the two models, we realize that in principle we can distinguish both KMM (minimal length) and ADV (maximum momentum) MCR via the Mandel's parameter since the former becomes more squeezed and exhibits sub-Poissonian distribution while the latter becomes “more classical” with super-Poissonian distribution. In\cite{deyfringCS}, it was shown that for the deformed GK coherent states, the KMM model predicts gravitational squeezing with sub-Poissonian distribution regardless of deformation $\beta$ and $\gamma$ (action variable). Similar behavior is observed in our results on both KMM-even (odd) cat states. However,\cite{GCSmaxP} showed that it is possible for ADV-deformed GK coherent states to exhibit both type of quantum statistics that depends on $\gamma$ factor. This generic behavior is in different to their cat states counterpart in our ADV models.

\subsection{Quadrature Squeezing of GKSCs}

Beside that the number squeezing effect introduced by the deformation of the GKSCs, here we consider another famous quantum optical effect, namely quadrature squeezing. Since the standard creation and annihilation operators ($A^{\dagger}$ and $A$) do not act as ladder operators in the new deformed Fock\rq{}s space\cite{bosso}, we search for modified ladder operators $A_q^{\dagger}$ and $A_q$ that act genuinely on the perturbed Hamiltonian eigenstates $|\xi^q_n\rangle$. The perturbative treatment is kept up to leading order in the deformation parameters. First, we define
\begin{eqnarray}
A_q|\xi^q_n\rangle= \sqrt{n}|\xi^q_{n-1}\rangle\ ;\hspace{1cm}A_q^\dagger|\xi^q_n\rangle= \sqrt{n+1}|\xi^q_{n+1}\rangle\ ;\hspace{1cm} N_q|\xi^q_n\rangle= n|\xi^q_{n}\rangle \label{Dladder}
\end{eqnarray}
where $q=(\beta,\alpha)$ refers to the class of deformations. Also, the modified ladder operators are required to obey the usual relations: $N_q=A_q^\dagger A_q\ ; \ A_q^\dagger = (A_q)^\dagger\ ; \ [A_q, A_q^\dagger]=1$. Consider KMM model, by a direct calculation we have
\begin{eqnarray}
\bigl(A_\beta - A\bigr)\bigl|\xi_{n}^{(ml)}\rangle 
&=& \sqrt{n}\bigl|\xi_{n-1}^{(ml)}\rangle - A \bigl|\xi_{n}^{(ml)}\rangle\nonumber\\
&=& \frac{\beta}{16}\left\{\Bigl(\sqrt{n\mathbf{P}[n, 4]}-\sqrt{(n+4)\mathbf{P}[n+1, 4]}\Bigr)|n+3\rangle\right.\nonumber\\
&&\hspace{0.65cm}\left.+\Bigl(\sqrt{(n-4)\mathbf{P}[n-3, 4]}-\sqrt{n\mathbf{P}[n-4, 4]}\Bigr)|n-5\rangle\right\}\nonumber\\
&=& -\frac{\beta}{4} \bigl(A^\dagger\bigr)^3|\xi_n^{(ml)}\rangle +O(\beta^2).
\end{eqnarray}
Thus, perturbatively up to leading order in $\beta$, the KMM-modified ladder operator is 
\begin{eqnarray}
A_\beta &=& A- \frac{\beta}{4} \bigl(A^\dagger\bigr)^3.
\end{eqnarray}
Similarly, for ADV model we have
\begin{eqnarray}
\bigl(A_\alpha - A\bigr) \bigl|\xi_{n}^{(mm)}\rangle &=& \sqrt{n} \bigl|\xi_{n-1}^{(mm)}\rangle - A \bigl|\xi_{n}^{(mm)}\rangle\nonumber\\
&=& \left[\frac{i\alpha}{\sqrt{8}} \bigl(A^\dagger\bigr)^2- 3A^2 - 3(2N+1)\right]|\xi_n^{(mm)}\rangle + O(\alpha^2)
\end{eqnarray}
and the perturbative expansion for ADV-modified ladder operator is given by
\begin{eqnarray}
A_\alpha &=& A +\frac{i\alpha}{\sqrt{8}} \bigl(A^\dagger\bigr)^2- 3A^2 - 3(2N+1)
\end{eqnarray}
where $N=A^\dagger A$ is the usual number operator.

Next, we define the two quadrature operators as
\begin{eqnarray}
&& X_q = \frac{1}{2}\Bigl(A_q + A_{q}^{\dagger}\Bigr)\hspace{0.35cm} ; \  \ Y_q= \frac{1}{2i}\Bigl(A_q - A_{q}^{\dagger}\Bigr)\\
\Rightarrow && X^2_q= \frac{1}{4}\Bigl(A^2_q + (A_{q}^{\dagger})^2 + 2 N_q + 1\Bigr)\ \ ; \ \ Y^2_q = -\frac{1}{4}\Bigl(A^2_q + (A_{q}^{\dagger})^2 - 2 N_q - 1\Bigr)\nonumber
\end{eqnarray}
and it is clear that $X_q$ and $Y_q$ are essentially dimensionless position and momentum operators
\begin{eqnarray}
X=\sqrt{\frac{2\hbar}{m\omega}}\ X_q\ ; \ P = \sqrt{2m\hbar\omega}\ Y_q . \nonumber
\end{eqnarray}

The square of the uncertainties in state $|\psi_{gksc}\rangle_q$ is
\begin{eqnarray}
\bigl(\Delta X_q\bigr)^2 &=& _q\langle\psi_{gksc}|X_q^2|\psi_{gksc}\rangle_q -\ _q\langle\psi_{gksc}|X_q|\psi_{gksc}\rangle^2_q\\
\bigl(\Delta Y_q\bigr)^2 &=& _q\langle\psi_{gksc}|Y_q^2|\psi_{gksc}\rangle_q -\ _q\langle\psi_{gksc}|Y_q|\psi_{gksc}\rangle^2_q .
\end{eqnarray}

We plot the quadrature $\bigl(\Delta X_q\bigr)^2$ and $\bigl(\Delta Y_q\bigr)^2$ of even GKSCs for KMM model (with various deformation parameters $\beta$) as the function of the average energy $J$ in Fig.(9a) and Fig.(9b) respectively. The odd GKSCs generally produce the similar trends.
\begin{figure}[h!]
    \centering
    \begin{subfigure}[b]{0.48\textwidth}
        \includegraphics[width=\textwidth]{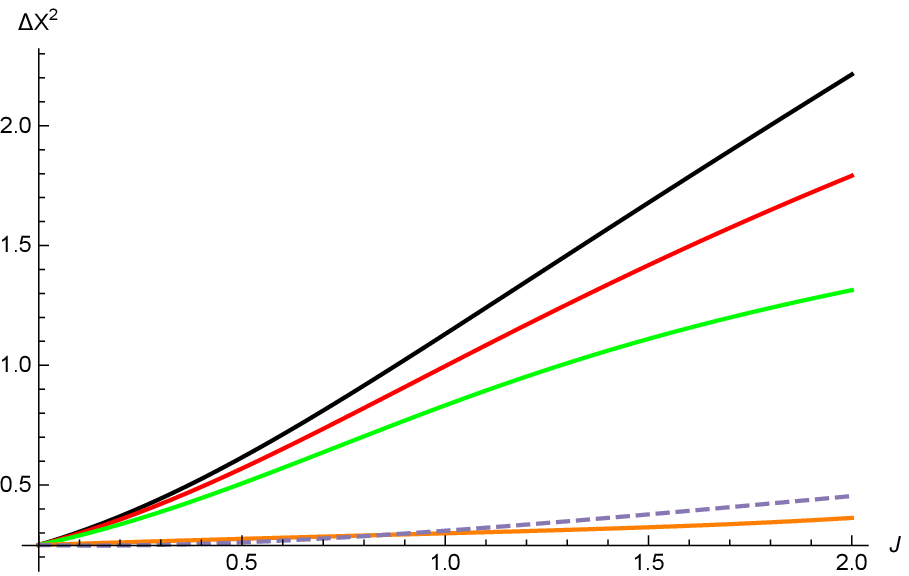}
        \caption{Quadrature $(\Delta X_\beta)^2$ for the even GKSCs.}
        \label{fig9a}
    \end{subfigure}
    ~ %add desired spacing between images, e. g. ~, \quad, \qquad, \hfill etc. 
    \begin{subfigure}[b]{0.48\textwidth}
        \includegraphics[width=\textwidth]{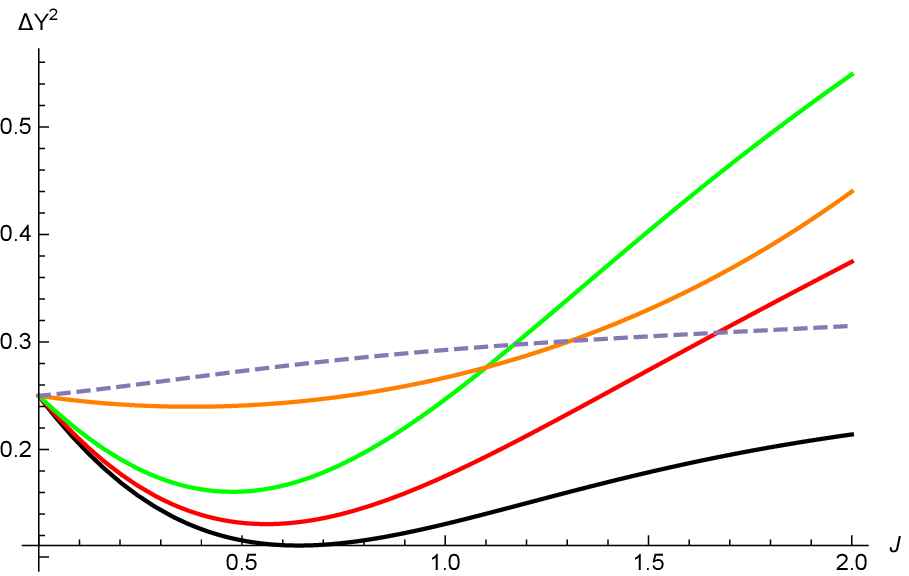}
        \caption{Quadrature $(\Delta Y_\beta)^2$ for the even GKSCs.}
        \label{fig9b}
    \end{subfigure}
    \caption{The quadrature $(\Delta X_\beta)^2$ and $(\Delta Y_\beta)^2$ for the even GKSCs with different values of deformation parameter $\beta$. Black line is the undeformed case while red, green, orange and dashed lines are the deformed cases with $\beta = (0.05, 0.1, 0.3, 1.2)$ respectively.}\label{fig: QuadratureKMM}
\end{figure}
We see that unlike the coherent state, uncertainties in two quadratures in deformed GKSCs are not equal to each other. The quadrature $(\Delta X_\beta)^2$ is squeezed below the standard case, i.e. GKSCs ($\beta=0$), whereas the quadrature $(\Delta Y_\beta)^2$ is expanded correspondingly. This means that we can reduce the quantum noise in $X_\beta$ variable. The condition eventually translates to more precise localization in position. Moreover, the total quantum noise $T:=(\Delta X_q)^2+(\Delta Y_q)^2$ also decreases when compared to the standard case. In Fig.(10a) and (10b), we plot the quantum noise in position $(\Delta X_\beta)^2$ and the total quantum noise $T$ as the function of $J$ and the deformation parameters $\beta$. We can adjust the values of $J$ with varying deformation parameters $\beta$, we manage to obtain the so-called ``ideal squeezed state", which is the minimum uncertainty states with smallest squeezing in $(\Delta X_\beta)^2$. Numerically, we obtain such a state with $\beta=1.00001$ and $J=0.658773$. It is clearly shown in Fig. (10a). Furthermore, interestingly we observe that for any fix energy $J$, the total quantum noise $T$ reduces for increasing $\beta$ up to a certain threshold $\beta_T$ and increases when $\beta>\beta_T$. At the same time, the total quantum noise is kept bounded from above by the standard ($\beta=0$) reading as clearly shown in Fig.(10b).   
\begin{figure}[h!]
    \centering
    \begin{subfigure}[b]{0.48\textwidth}
        \includegraphics[width=\textwidth]{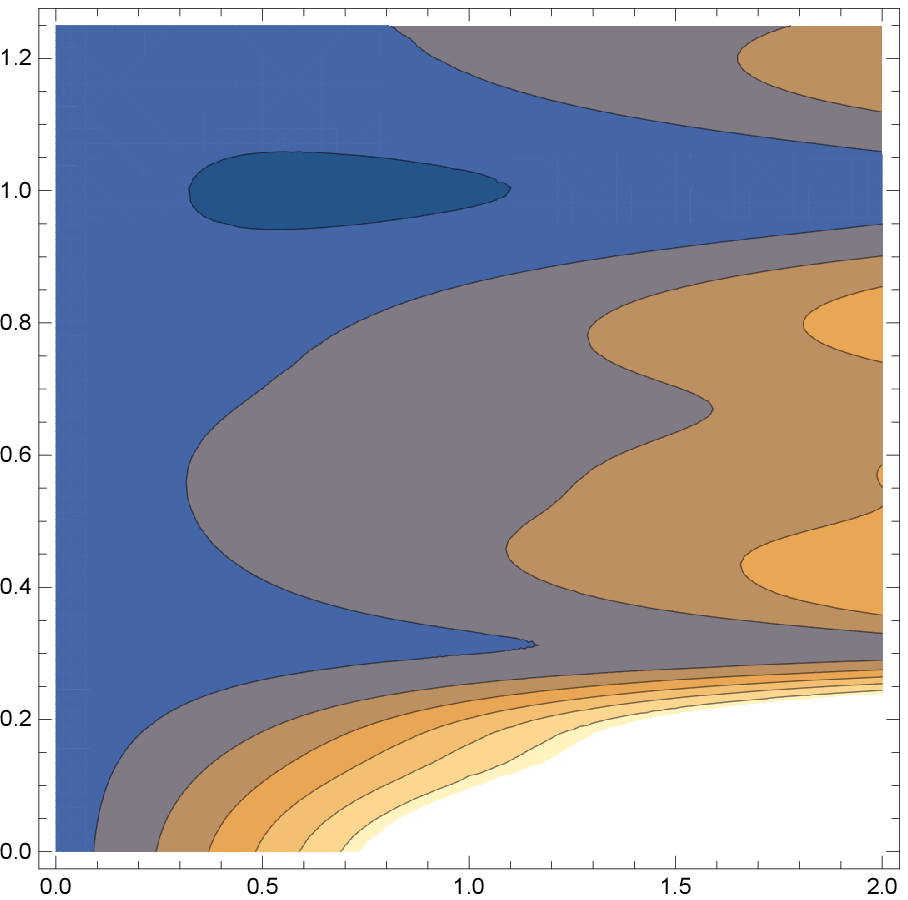}
        \caption{Quantum noise $(\Delta X_\beta)^2$ for the even GKSCs as function of $J$ and $\beta$.}
        \label{fig10a}
    \end{subfigure}
    ~ %add desired spacing between images, e. g. ~, \quad, \qquad, \hfill etc. 
    \begin{subfigure}[b]{0.48\textwidth}
        \includegraphics[width=\textwidth]{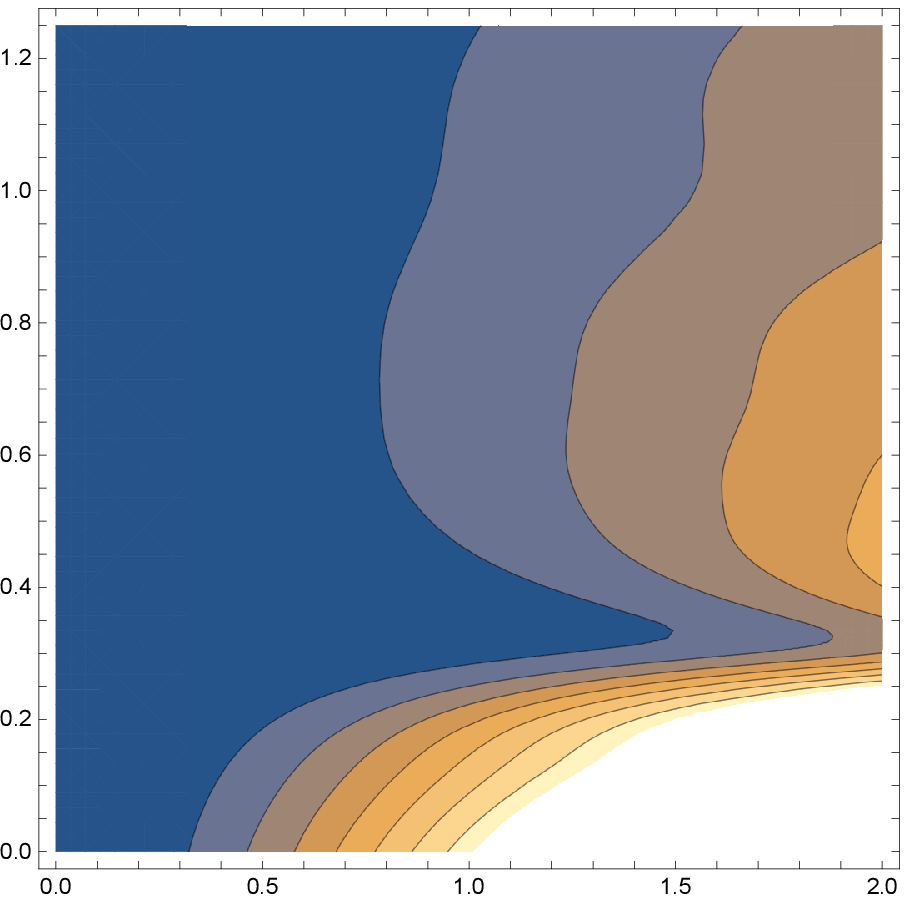}
        \caption{Total quantum noise $T=(\Delta X_\beta)^2 + (\Delta Y_\beta)^2$ for the even GKSCs as function of $J$ and $\beta$.}
        \label{fig10b}
    \end{subfigure}
    \caption{Contour plot of the quantum noise as the function of $J$ (horizontal axis) and deformation parameters $\beta$ (vertical axis). Darker colors refer to smaller values.}\label{fig: Quantnoise}
\end{figure}

Next, we consider the generalized uncertainty of the quadrature pairs $(\Delta X_\beta)(\Delta Y_\beta)$. 
\begin{figure}[h!]
    \centering
        \includegraphics[width=10cm]{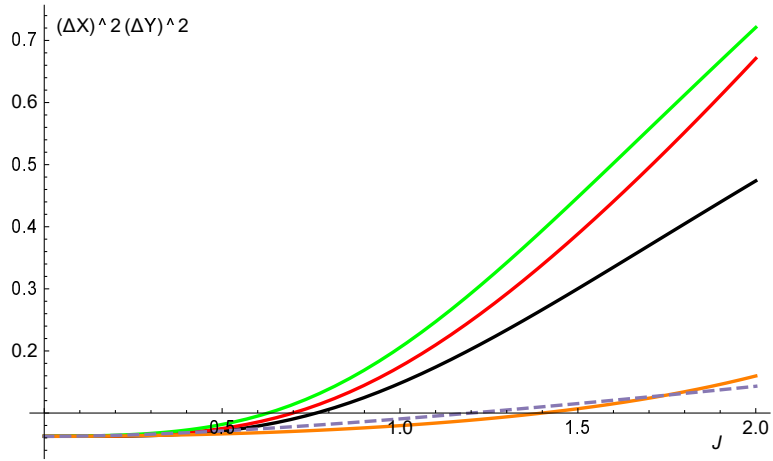}
        \caption{Generalized uncertainty $(\Delta X_\beta)^2 (\Delta Y_\beta)^2$ for the even GKSCs as function of $J$ with different values of deformation parameter $\beta$. Black line is the undeformed case while red, green, orange and dashed lines are the deformed cases with $\beta = (0.05, 0.1, 0.3, 1.2)$ respectively.}
        \label{fig11}
\end{figure}
In Fig.(11), we see that there are constraints on both $J$ and deformation parameters $\beta$ in order to satisfy the inequality
\begin{eqnarray}
(\Delta X_\beta)\cdot(\Delta Y_\beta)\geq (\Delta X_{\beta=0})\cdot(\Delta Y_{\beta=0}). \label{gup1}
\end{eqnarray}
From the form of MCR in \eqref{mcr1}, we do not expect KMM deformed-GKSCs to approach classical phase (i.e. with vanishing GUP) and hence the inequality \eqref{gup1} has to be strictly satisfied. Numerically, deformation is meaningful if $\beta\leq\beta_{max}\approx 0.23$ for the range of energy $J$ we considered. However, in higher energy sector which is physically more preferred from phenomenological point of view, it remains possible to satisfy \eqref{gup1} with larger values of $\beta$. Similar results are obtained for the GKSCs odd state.

For ADV model, we plot the quadrature $(\Delta X_\alpha)^2$ and $(\Delta Y_\alpha)^2$ of even GKSCs as the function of the average energy $J$ in Fig.(12a) and Fig.(12b) respectively. 
\begin{figure}[h!]
    \centering
    \begin{subfigure}[b]{0.48\textwidth}
        \includegraphics[width=\textwidth]{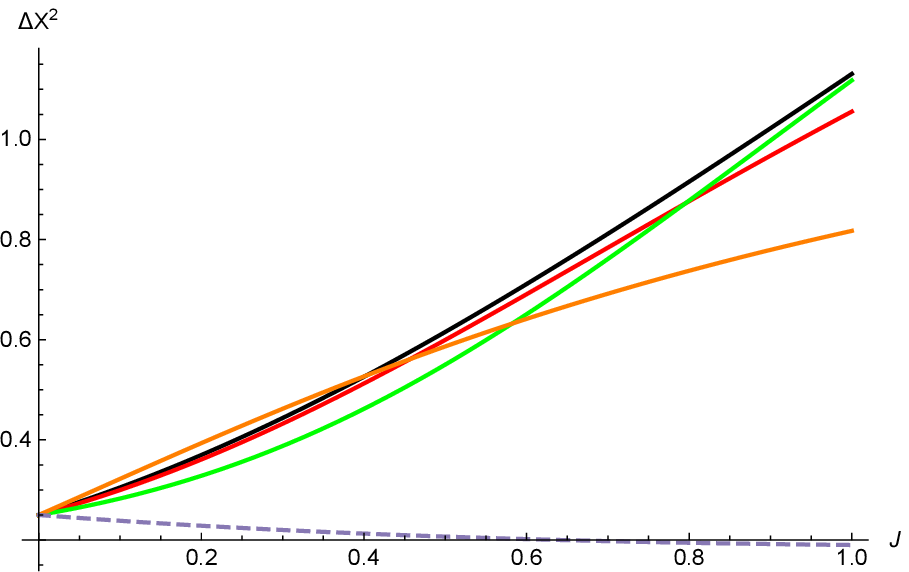}
        \caption{Quadrature $(\Delta X_\alpha)^2$ for the even GKSCs.}
        \label{fig12a}
    \end{subfigure}
    ~ %add desired spacing between images, e. g. ~, \quad, \qquad, \hfill etc. 
    \begin{subfigure}[b]{0.48\textwidth}
        \includegraphics[width=\textwidth]{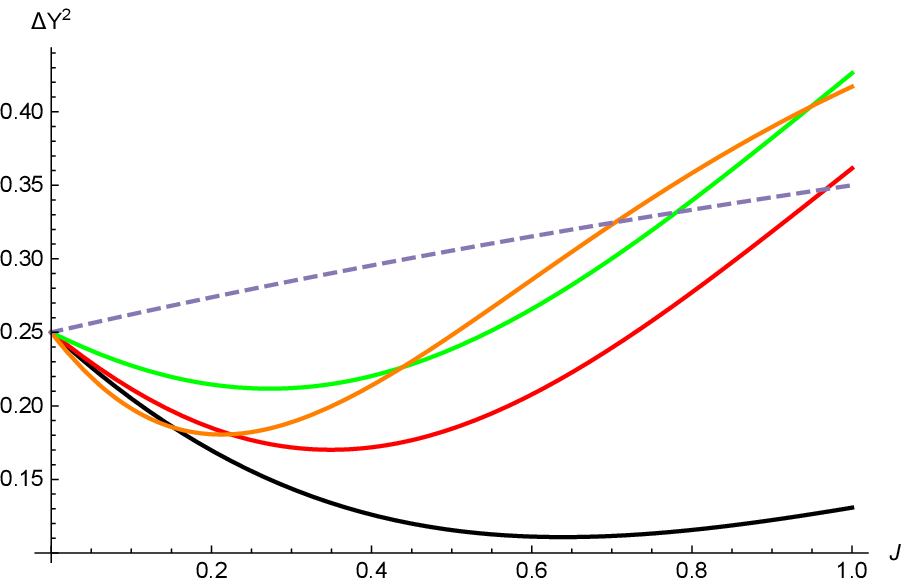}
        \caption{Quadrature $(\Delta Y_\alpha)^2$ for the even GKSCs.}
        \label{fig12b}
    \end{subfigure}
    \caption{The quadrature $(\Delta X_\alpha)^2$ and $(\Delta Y_\alpha)^2$ for the even GKSCs with different values of deformation parameter $\alpha$. Black line is the undeformed case while red, green, orange and dashed lines are the deformed cases with $\alpha = (0.22, 0.35, 0.55, 1.00)$ respectively.}\label{fig: QuadratureADV}
\end{figure}
Unlike KMM model, deformed quantum noise in first quadrature $(\Delta X_\alpha)^2$ can be increased or decreased (depends on the range of energy $J$ and the amount of deformation) as compared to the undeformed case. In contrast, $(\Delta X_\beta)^2$ always decreases while $(\Delta Y_\beta)^2$ always increases in KMM model. We infer that improvement in spatial resolution and thus reduction of quantum noise are not guaranteed for ADV-deformed GKSCs. Next, we consider the generalized uncertainty principle $(\Delta X_\alpha)(\Delta Y_\alpha)$ in the Fig.[13].
\begin{figure}[h!]
    \centering
        \includegraphics[width=10cm]{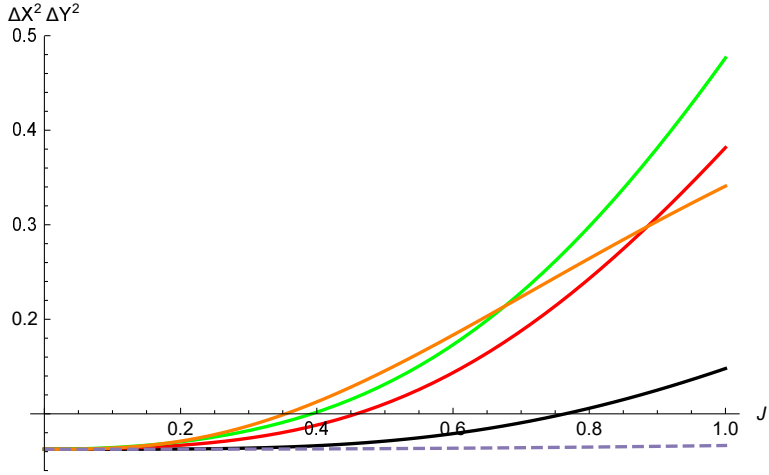}
        \caption{Generalized uncertainty $(\Delta X_\alpha)^2 (\Delta Y_\alpha)^2$ for the even GKSCs as function of $J$ with different values of deformation parameter $\alpha$. Black line is the undeformed case while red, green, orange and dashed lines are the deformed cases with $\alpha = (0.22, 0.35, 0.55, 1.00)$ respectively.}
        \label{fig13}
\end{figure}
In contrast to KMM model, we do not required the quadratures in ADV model to strictly satisfy \eqref{gup1}. From \eqref{mcr2}, we expect the emergence of classicality in ADV model when the energy scale in approaching maximum momentum $P \rightarrow P_{max} \approx \frac{1}{\widetilde{\alpha}}$. In summary, we observed that the two classes of quantum deformation induce slight differences in their effects on the GKSCs.

\section{Husimi Distribution}

It is a well known fact that SCs are formed by a superposition of two macroscopically distinguishable states was emphasized by Schr$\ddot{\text{o}}$dinger himself in his original work. To study the superposition effect in deformed-GKSCs, we consider the phase space distribution as a natural quantifier \cite{dajka,klauderB,scully,mandel}. In literature, there are two important probability distributions to characterize the phase space properties, the so-called Wigner $\mathcal{W}$-distribution and Husimi $\mathcal{Q}$-distribution. Husimi distribution is given by the coherent state expectation value of the density operator or equivalently the overlap between the wave function and coherent state. It is strictly nonnegative by construction. Here, we choose to explore the Husimi distribution\footnote{Note that this choice would be clearly insufficient if we are interested in nonclassical properties indicated by the deformed GKSCs. Nonclassicality can be well understand in a negative value of the Wigner $\mathcal{W}$-function\cite{dajka,kenfack,case}.} for the KMM/ADV-deformed GKSCs.

For the GKSCs, since the density of matrix is $\rho_q=|\psi_{gksc}\rangle_q \bigotimes \phantom{|}_q\langle \psi_{gksc}|$, we define the corresponding Husimi $\mathcal{Q}$-function as
\begin{eqnarray}
\mathcal{Q} (z):=\frac{1}{\pi} \langle z|\rho_q|z\rangle =\frac{1}{\pi} |\langle z|\psi_{gksc}\rangle_q|^2.
\end{eqnarray}
This definition ensures the Husimi function to be normalized $\int dz^2 \mathcal{Q}(z)=1$ and bounded $0<\mathcal{Q}(z)<1/\pi$. We have the explicit form in Fock\rq{}s basis as
\begin{eqnarray}
\mathcal{Q}(z)=\frac{1}{\pi} e^{-|z|^2} \sum_{n,m} C_n C^{\ast}_m \frac{z^m (z^\ast)^n}{\sqrt{n!m!}}
\end{eqnarray}
where the constant $C_n$ is
\begin{eqnarray}
C_n = N^q_{gksc} \frac{J^{n/2}}{\sqrt{\rho^q_n}} e^{-i\epsilon^q_n\gamma} \bigl[1+ \cos (\phi-\epsilon^q_n \pi)\bigr] .
\end{eqnarray}

By setting $J = 10$; $\phi = 0 = \gamma, q=0$, we illustrate the Husimi distribution for even $\phi = 0$ standard GKSCs in the following figures. In Fig.(14), two sharp peaks corresponding to two coherent states in standard SCs is clearly shown for the underformed case.
\begin{figure}[h!]
\begin{center}
\epsfig{file=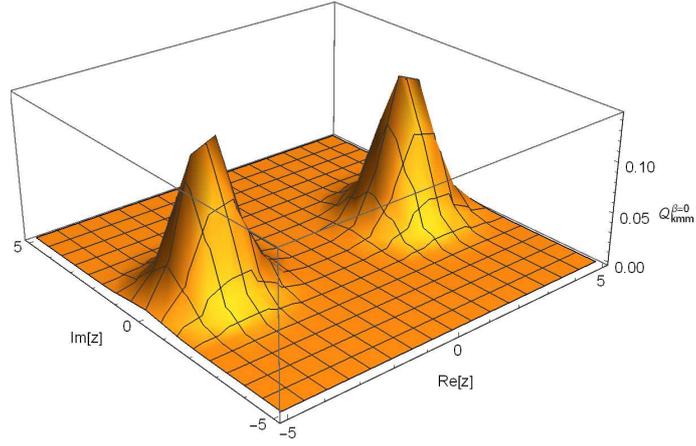, width=9cm}
\caption{The Husimi function $\mathcal{Q}(z)$ for the undeformed SCs case with $J = 10$, $\phi = 0 = \gamma$.}
\label{fig14}
\end{center}
\end{figure}

For the KMM model, when the deformation parameter $\beta$ increases, the well localized cat-like state gradually merges together and the two coherent states become indistinguishable. This is illustrated in Fig.(15a) and Fig.(15b).

\begin{figure}[h!]
    \centering
    \begin{subfigure}[b]{0.48\textwidth}
        \includegraphics[width=\textwidth]{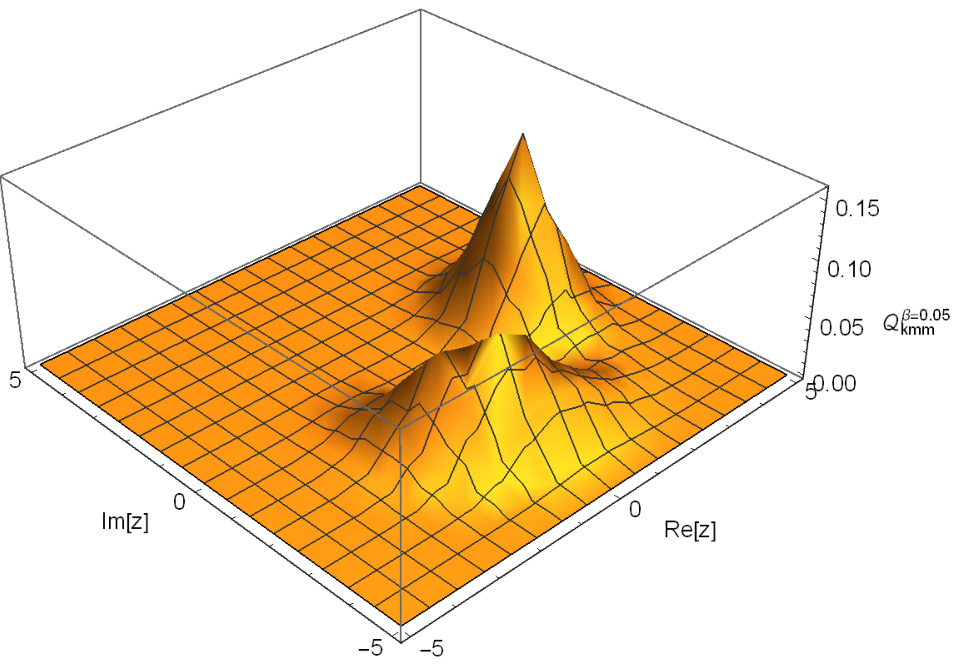}
        \caption{$\beta=0.05$}
        \label{fig15a}
    \end{subfigure}
    ~ %add desired spacing between images, e. g. ~, \quad, \qquad, \hfill etc. 
    \begin{subfigure}[b]{0.48\textwidth}
        \includegraphics[width=\textwidth]{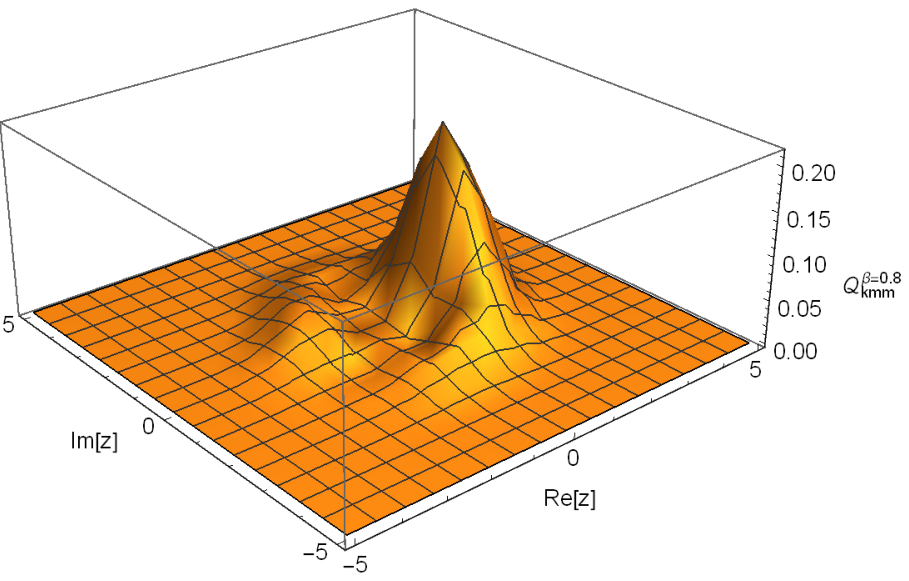}
        \caption{$\beta=0.8$}
        \label{fig15b}
    \end{subfigure}
    \caption{The Husimi function $\mathcal{Q}(z)$ for the KMM-deformed GKSCs with different deformation $\beta$.}
		\label{fig: HF_KMM}
\end{figure}

For ADV model, the similar results are obtained as clearly shown in Fig.(16a) and Fig.(16b). Although the peaks are clearly observed for small value of $\alpha$, as $\alpha$ increases, the peaks become nonseparable. The result for odd cat states $\phi = \pi$ are essentially similar to the even cat states. We interpret this result as some kind of gravitational decoherence \cite{bassi,pikovskizych} due to minimal length/maximum momentum. In an addition, for a deformed-GKSCs with ``peaks merged" Husimi function, it does not seem to have an effective method to determine which form of the deformation does the system undergoes.

\begin{figure}[h!]
    \centering
    \begin{subfigure}[b]{0.48\textwidth}
        \includegraphics[width=\textwidth]{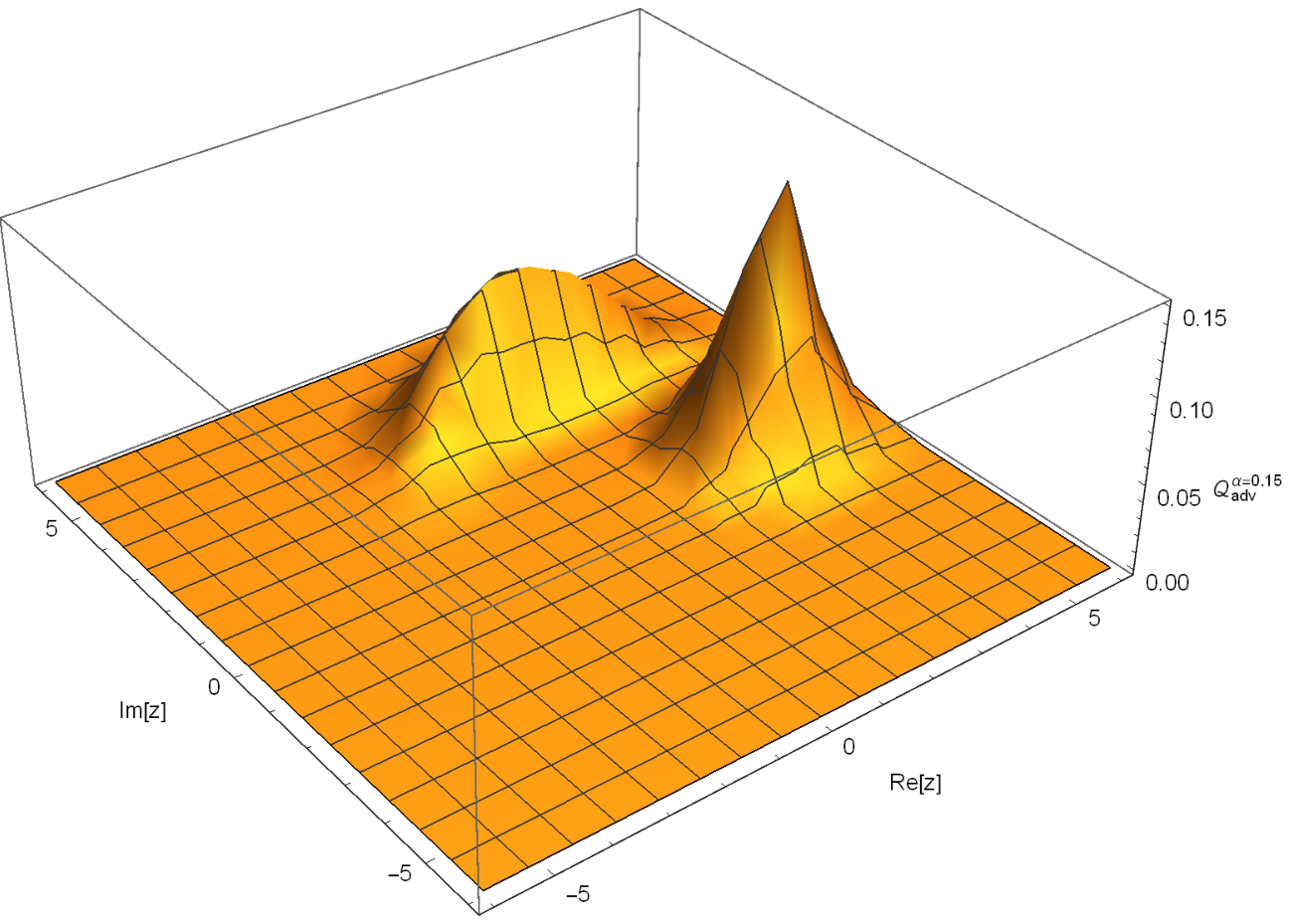}
        \caption{$\alpha=0.15$}
        \label{fig16a}
    \end{subfigure}
    ~ %add desired spacing between images, e. g. ~, \quad, \qquad, \hfill etc. 
    \begin{subfigure}[b]{0.48\textwidth}
        \includegraphics[width=\textwidth]{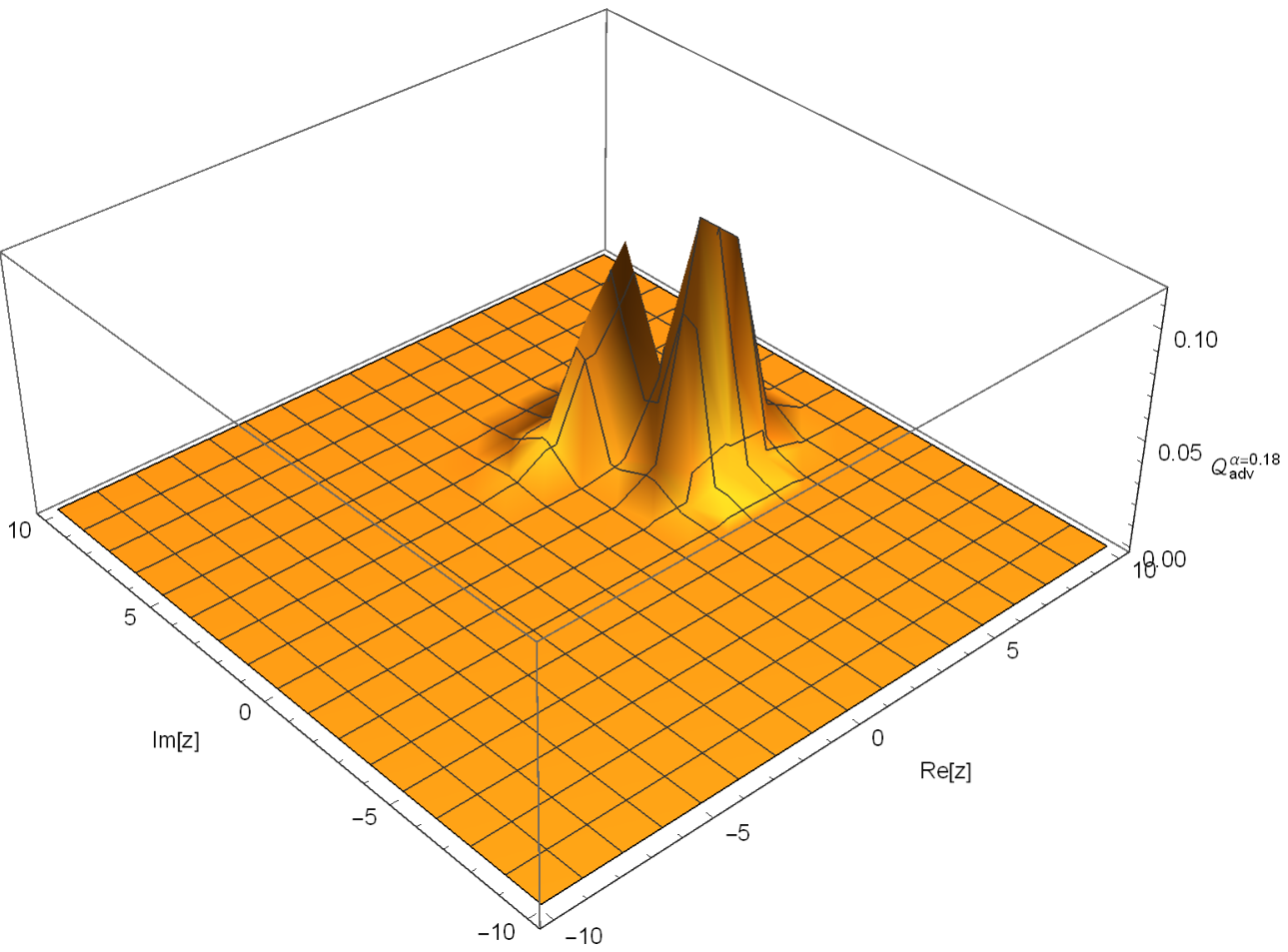}
        \caption{$\alpha=0.18$}
        \label{fig16b}
    \end{subfigure}
    \caption{The Husimi function $\mathcal{Q}(z)$ for the ADV-deformed GKSCs with different deformation $\alpha$.}
		\label{fig: HF_ADV}
\end{figure}

\section{Conclusion}

In this paper, we have successfully constructed deformed Gazeau-Klauder Schr$\ddot{\text{o}}$dinger cat states (GKSCs) under two important quantum gravity phenomenological models which exhibit minimal length scale (KMM model) and maximum momentum scale (ADV model). Firstly we review the generalized Heisenberg algebra scheme. This allows us to define the characteristic function and the weight function of the deformed-Gazeau-Klauder coherent states (GKCs). Formally, we can proceed to take the linear superposition of such states to obtain the deformed-GKSCs.

We have computed the probability density and entropy distribution function in both models (for even and odd cat states) and subsequently studied their behavior in terms of the deformation parameters. Under finite deformation, even GKSCs possess small, nonvanishing odd probability distribution while odd GKSCs possess nonvanishing even probability distribution. We interpret this deviation from standard GKSCs as gravitational induced mixture of the two probability distributions to detect odd and even GKSCs respectively. The entropy of the system increases in the presence of deformation. These observations are valid in the both models and is insensitive to the type of deformation models.

To study the nonclassical behavior, we have considered number squeezing (defined in terms of Mandel's $Q$ parameter) and quadrature squeezing. We realized that in principle we can distinguish both KMM and ADV model via the Mandel's parameter since the former becomes more squeezed and exhibits sub-Poissonian distribution while the latter becomes “more classical” with super-Poissonian distribution. Furthermore, first quadrature squeezing $\Delta X_\beta$ in KMM model always reduces with increasing deformation parameter $\beta$. This also means that KMM deformation leads to better spatial resolution and reduction of quantum noise. However, this desired behavior is not guaranteed in ADV model since $\Delta X_\alpha$ can be larger or smaller than the undeformed case. Lastly, we observed that the quantum coherency of deformed-GKSCs is destroyed by the deformations as clearly shown in the indistinguishability of the peaks in the Husimi function $\mathcal{Q}(z)$.

\section{Acknowledgment}

We thank the referee(s) for the valuable and critical comments for improvement of the presentation. 

%\section*{Appendix A:}


\newpage
\begin{thebibliography}{99}

%Abstract
\bibitem{subirCS} 
S. Ghosh and P. Roy, Phys. Lett. B {\bf 711}, 423 (2012).
%GCS GUP

\bibitem{deyfringCS} 
S. Dey, A. Fring, Phys. Rev. D {\bf 86}, 064038 (2012).
%GCS minimal length

\bibitem{pedramCS} 
P. Pedram, Int. J. Mod. Phys. D {\bf 22}, 1350004 (2013).
%pedram coherent state

\bibitem{GCSmaxP} 
C.L. Ching and W.K. Ng, Phys. Rev. D {\bf 88}, 084009 (2013).
%GCS maximum momentum

\bibitem{dajka} 
J. Dajka and J. Luczka, J. Phys. A: Math. Theo. {\bf 45}, 244006 (2012). 



%Introduction

\bibitem{polchinski} 
J. Polchinski, String Theory Vol.1 and 2 (Cambridge University Press, UK, 2005).

\bibitem{becker} 
K. Becker, String Theory and M-Theory: A Modern Introduction (Cambridge University Press, UK, 2007).

\bibitem{rovelli} 
C. Rovelli, Living Rev. Relativity, {\bf 11}, 5 (2008).
%review LQG

\bibitem{ashtekar} 
A. Ashtekar and J. Pullin, Loop Quantum Gravity: The First 30 Years (World Scientific, Singapore, 2017).
%review LQG

\bibitem{connes} 
A. Connes, Noncommutative Geometry (Academic Press, Sand Diego CA 1994).
%review noncommutative geometry

\bibitem{douglas} 
M.R. Douglas and N.A. Nekrasov, Rev. Mod. Phys. {\bf 73}, 977 (2001).
%review noncommutative field theory

\bibitem{reuter} 
M. Reuter and F. Saueressig, Phys. Rev. D {\bf 65}, 065016 (2002). 
%RG flow to gravity

\bibitem{adler} 
R.J. Adler, Am. J. Phys {\bf 78}, 925 (2010).
%Quantized Space-Time

\bibitem{snyder} 
H.S. Snyder, Phys. Rev. {\bf 71}, 38 (1947).
%Quantized Space-Time

\bibitem{kempf1} 
A. Kempf, G. Mangano and R.B. Mann, Phys. Rev. D {\bf 52}, 1108 (1995).
%GUP original

\bibitem{kempf2} 
A. Kempf, J. Phys. A {\bf 30}, 2093 (1997).
%GUP original

\bibitem{laynam1} 
L.N. Chang, D. Minic, N. Okamura and T. Takeuchi, Phys. Rev. D {\bf 65}, 125028 (2002).

\bibitem{lewis} 
Z. Lewis and T. Takeuchi, Phys. Rev. D {\bf 84}, 105029 (2011). 

\bibitem{laynam2}
L.N. Chang, Z. Lewis, D. Minic and T. Takeuchi, Advances in HEP, 493514 (2011).

\bibitem{quesne1}
C. Quesne and V.M. Tkachuk, Phys. Rev. A {\bf 81}, 012106 (2010).

\bibitem{brau}
F. Brau, J. Phys. A {\bf 32}, 7691 (1999).
%quadratic phenom

\bibitem{harbach} 
U. Harbach, S. Hossenfelder, M. Bleicher and H. Stocker, Phys. Lett. B {\bf 584}, 109 (2004).
%quadratic phenom

\bibitem{dasvagenas} 
S. Das and E.C. Vagenas, Phys. Rev. Lett. {\bf 101}, 221301 (2008). 
%quadratic phenom

\bibitem{nozari} 
K. Nozari and P. Pedram, Euro. Phys. Lett. {\bf 92}, 50013 (2010).
%quadratic phenom

\bibitem{bouaziz} 
D. Bouaziz and N. Ferkous,  Phys. Rev. A {\bf 82}, 022105 (2010). 
%quadratic phenom

\bibitem{kober} 
M. Kober, Phys. Rev. D {\bf 82}, 085017 (2010).
%quadratic phenom

\bibitem{pedram1}
P. Pedram, J. Phys. A: Math. Theor. {\bf 45}, 505304 (2012). 
%quadratic phenom

\bibitem{sabine} 
S. Hossenfelder, Living Rev. Relativity, {\bf 16}, 2 (2013).
%review

\bibitem{tawfik} 
A.N. Tawfik and A.M. Diab, Rept.Prog.Phys, {\bf 78}, 126001 (2015).
%gupreview

\bibitem{magueijo1} 
J. Magueijo and L. Smolin, Phys. Rev. Lett. {\bf 88}, 190403 (2002).

\bibitem{magueijo2} 
J. Magueijo and L. Smolin, Phys. Rev. D {\bf 67}, 044017 (2003).

\bibitem{cortes} 
J.L. Cortes and J. Gamboa, Phys. Rev. D {\bf 71}, 065015 (2005).

\bibitem{camelia}
G. Amelino-Camelia, Phys. Lett. B {\bf 510}, 255 (2001).

\bibitem{das1} 
A.F. Ali, S. Das and E.C. Vagenas, Phys. Lett. B {\bf 678}, 497 (2009).

\bibitem{das2} 
A.F. Ali, S. Das and E.C. Vagenas, Phys. Rev. D {\bf 84}, 044013 (2011).
%details, jocobi etc

\bibitem{das3} 
S. Das, E.C. Vagenas and A.F. Ali, Phys. Lett. B {\bf 690}, 407 (2010).

\bibitem{pedram2}
P. Pedram, Phys. Lett. B {\bf 702}, 295 (2011).

\bibitem{mignemi} 
S. Mignemi, Phys. Rev. D {\bf 84}, 025021 (2011).
%dsr antisynder

\bibitem{jizba} 
P. Jizba, H. Kleinert and F. Scardigli, Phys. Rev. D {\bf 81}, 084030 (2010).

\bibitem{maggiore} 
M. Maggiore, Phys. Lett. B {\bf 319}, 83 (1993).
%cutoff

\bibitem{battisti} 
M.V. Battisti, Phys. Rev. D {\bf 79}, 083506 (2009).
%cutoff

\bibitem{pedram3} 
P. Pedram, Phys. Lett. B {\bf 714}, 317 (2012); {\bf 718}, 638 (2012).
%higher order GUP. max mom

\bibitem{chingspectra} 
C.L. Ching, R. Parwani, and K. Singh, Phys. Rev. D {\bf 86}, 084053 (2012).
%maximum momentum energy spectra

\bibitem{camelia-dsr} 
G. Amelino-Camelia, Int. J. Mod. Phys. D {\bf 11}, 35 (2002); {\bf 11}, 1643 (2002).

\bibitem{magueijo-dsr} 
J. Magueijo and L. Smolin, Phys. Rev. Lett. {\bf 88}, 190403 (2002).

\bibitem{pikovski} 
I. Pikovski, M.R. Vanner, M. Aspelmeyer, M. Kim and C. Brukner, Nature Physics {\bf 8}, 393 (2012).
%experiment on GUP

\bibitem{marin} 
F. Marin, F. Marino and \textit{et. al}, Nature Physics {\bf 9}, 71 (2012).
%\textit{Gravitational bar detectors set limits to Planck-scale physics on macroscopic variables}, %experiment on GUP

\bibitem{bawaj}
M. Bawaj, C. Biancofiore and \textit{et. al}, Nature Communications {\bf 6}, 7503 (2015).
%\textit{Probing deformed commutators with macroscopic harmonic oscillators} %experiment on GUP

\bibitem{gao} 
D. Gao and M. Zhan, Phys. Rev. A {\bf 94}, 013607 (2016).
%experiment on GUP

\bibitem{rossi} 
M.A.C. Rossi, T. Giani and M.G.A Paris, Phys. Rev. D {\bf 94}, 024014 (2016).
%experiment on GUP

\bibitem{deyopto} 
S. Dey et al., Nuclear Physics B {\bf 924(C)}, 578 (2017).
%Probing noncommutative theories with quantum optical experiments

\bibitem{howl} 
R. Howl, L. Hackermüller, D.E. Bruschi and I. Fuentes, Advances in Physics: X, {\bf 3 (1)}, 1383184 (2018).

\bibitem{klauder0} 
J.R. Klauder, Ann. Phys. {\bf 11}, 123 (1960).
%CS original

\bibitem{klauder1} 
J.R. Klauder, J. Math. Phys. {\bf 4}, 1055 (1963).
%CS original

\bibitem{glauber1} 
R.J. Glauber, Phys. Rev. Lett. {\bf 10}, 84 (1963).
%CS original

\bibitem{sudarshan1}
E.C.G. Sudarshan, Phys. Rev. Lett. {\bf 10}, 277 (1963).
%CS original

\bibitem{gazeau1}
J.P. Gazeau, Coherent States in Quantum Physics (Wiley Press, Berlin, 2009).

\bibitem{klauderB} 
J.R. Klauder and B.S. Skagertan, Coherent States: Applications in Physics and Mathematical Physics  (World Scientific, Singapore, 1985).
%GK original

\bibitem{scully}
M.O. Scully and M.S. Zubairy, Quantum Optics (Cambridge University Press, UK, 1997).

\bibitem{mandel} 
L. Mandel and E. Wolf, Optical Coherence and Quantum Optics (Cambridge University Press, UK, 1995).

\bibitem{walls} 
D.F. Walls, Nature (London) {\bf 306}, 141 (1983).
% Squeezed states of light

\bibitem{loudon} 
R. Loudon and P.L. Knight, J. Mod. Opt. {\bf 34}, 709 (1987).
% Squeezed light

\bibitem{angelova} 
M. Angelova, A. Hertz and V. Hussin, J.Phys. A: Math. Theo. {\bf 45}, 244007 (2012).

\bibitem{kok1}
P. Kok, W.J. Munro, K. Nemoto, T.C. Ralph, J.P. Dowling and G.J. Milburn, Rev. Mod. Phys. {\bf 79}, 135 (2007).
%Linear optical quantum computing with photonic qubits

\bibitem{kok2} 
P. Kok and B.W. Lovett , Optical Quantum Information Processing (Cambridge University Press, UK, 2010).

\bibitem{yamamoto}
Y. Yamamoto and H.A. Haus, Rev. Mod. Phys. {\bf 58}, 1001 (1986).
%Preparation, measurement and information capacity of optical quantum states

\bibitem{giovannetti}
V. Giovannetti, S. Lloyd and L. Maccone, Nature Photonic {\bf 5}, 222 (2011).
%Advances in quantum metrology

\bibitem{hillery}
M. Hillery, Phys. Rev. A {\bf 61}, 022309 (2000).
%Quantum cryptography with squeezed states

\bibitem{anisimov}
P.M. Anisimov, G.M. Raterman, A. Chiruvelli, W.N. Plick, S.D. Huver, H. Lee and J.P. Dowling, Phys. Rev. Lett. {\bf 104}, 103602 (2010).
%Quantum Metrology with Two-Mode Squeezed Vacuum: Parity Detection Beats the Heisenberg Limit

\bibitem{aasi}
J. Aasi, et al., Nature Photonics {\bf 7}, 613 (2013).
%Enhanced sensitivity of the LIGO gravitational wave detector by using squeezed states of light

\bibitem{deyq} 
S. Dey, Phys. Rev. D {\bf 91} 044024 (2015).
%q-deformed cat state

\bibitem{deyfringR} 
S. Dey, A. Fring and V. Hussin, Springer Proc. Phys. {\bf 205}, 209 (2018).
%review on squeezed state with MCR

\bibitem{deyhussin} 
S. Dey and V. Hussin, Phys. Rev. D {\bf 91} 124017 (2015).
% entangled squeezed state

\bibitem{deyzelaya} 
K. Zelaya, S. Dey and V. Hussin, Phys. Lett. A {\bf 382}, 3369 (2018).
%generalized squeezed states


\bibitem{curado} 
E.M.F. Curado, M.A. Rego-Monteiro and Ligia M.C.S. Rodrigues, Phys. Rev. A {\bf 87} 052120 (2013).
%GHA with decoherence

\bibitem{berrada} 
K. Berrada and H. Eleuch, Phys. Rev. D {\bf 100} 016020 (2019).
%noncommutative deformed cat state with decoherence

\bibitem{yingwu} 
Y. Wu and X. Yang, Phys. Rev. D {\bf 73}, 067701 (2006).
%schrodinger cat noncommutative space

\bibitem{gha1} 
C. Quesne and N. Vansteenkiste, J. Phys. A {\bf 28}, 7019 (1995).
%generalized heisenberg algebra

\bibitem{gha2}
E.M.F. Curado and M.A. Rego-Monteiro, J. Phys. A {\bf 34}, 3253 (2001).
%generalized heisenberg algebra

\bibitem{gha3}
Y. Hassouni, E.M.F. Curado and M.A. Rego-Monteiro, Phys. Rev. A {\bf 71}, 022104 (2005).
%generalized heisenberg algebra

\bibitem{gazeau2}
J.P. Antoine, J.P. Gazeau, P. Monceau, J.R. Klauder and K.A. Penson, J. Math. Physics. {\bf 42}, 2349 (2001).
%GK orignal

\bibitem{klauder2} 
J.R. Klauder, Annals of Physics. {\bf 237}, 147 (1995).
%GK original

\bibitem{klauder3} 
J.R. Klauder, J.-P. Gazeau, J. Phys. A. {\bf 32}, 123 (1999).
%GK original


%Sect.(2)
\bibitem{schrodinger} 
E. Schrodinger, Naturwiss {\bf 14}, 664 (1926).
%CS orignal

\bibitem{csapp1} 
W.M. Zhang, D.H. Feng and R.Gilmore, Rev. Mod. Phys. {\bf 62}, 867 (1990).
%application of CS

\bibitem{csapp2} 
G.S. Agarwal, J. Benerji, Phys. Rev. A {\bf 64}, 023815 (2001). 
%application of CS

\bibitem{csapp3}
C. Quesne, J. Phys. A {\bf 35}, 9213 (2002).
%application of CS

\bibitem{gilm} 
V.G. Drinfeld, in: Proceedings of the International Congress of Mathematicians, Berkeley, California, 1986, Amer. Math. Soc., Providence, 1987, p. 798.
%group theoretical to CS

\bibitem{wigner} 
E. Wigner, Phys. Rev. {\bf 77}, 711 (1950).
%WHA (original)

\bibitem{yang} 
L.M. Yang, Phys. Rev. {\bf 84}, 788 (1951).
%WHA (original)

\bibitem{greenberg} 
0. W. Greenberg, Phys. Rev. Lett. {\bf 13}, 598 (1964). 
%WHA on QCD

\bibitem{plyushchay1} 
M.S. Plyushchay, Phys. Lett. B {\bf 320}, 91 (1994). 
%WHA on parastatistic

\bibitem{plyushchay2} 
M. S. Plyushchay, Nuclear Physics B {\bf 491}, 619 (1997).
%WHA on parastatistic

\bibitem{plyushchay3} 
M.S. Plyushchay, Int. J. of Mod. Phys. A {\bf 15}, 3679 (2000).
%WHA on parastatistic

\bibitem{plyushchay4} 
P.A. Horv$\acute{a}$thy and M.S. Plyushchay, Phys. Lett. B {\bf 595}, 547 (2004). 
%WHA on anyon

\bibitem{plyushchay5} 
P.A. Horv$\acute{a}$thy, M.S. Plyushchay and M. Valenzuela, Ann. Phys. {\bf 325}, 1931 (2010).
%WHA on anyons

\bibitem{plyushchay6} 
M.S. Plyushchay, Ann. Phys. {\bf 245}, 339 (1996).
%WHA on anyons


\bibitem{deghani1} 
A. Dehghani, B. Mojaveri, S. Shirin and M. Saedi, Ann. Phys. {\bf 362}, 659 (2015).
%WHA on Cat state

\bibitem{deghani2} 
A. Dehghani, B. Mojaveri and S.A. Faseghandis, Mod. Phys. Lett. A {\bf 34}, 1950104 (2019).
%WHA on Cat state

\bibitem{dehdashti} 
Sh. Dehdashti, M.B. Harouni, A. Mahdifar and R. Roknizadeh, Laser Phys. {\bf 24}, 055203 (2014).
%WHA on Cat state

\bibitem{deghani3} 
A. Dehghani, B. Mojaveri, R.J. Bahrbeig, F. Nosrati and R.L. Franco, Journal of the Optical Society of America B {\bf 36 (7)}, 1858 (2019).
%WHA on Cat state

%Sect.(3)

\bibitem{kempf3} 
A. Kempf, J. Math. Phys. (NY){\bf 35}, 4483 (1994).

\bibitem{bagchi} 
B. Bagchi and A. Fring, Phys. Lett. A {\bf 373}, 4307 (2009).


\bibitem{bonneau}
G. Bonneau, J. Faraut and G. Valent, Am. J. Phys. {\bf 69}(3), 322 (2001).

\bibitem{araujo}
V.S. Araujo, F.A.B. Coutinho and J.F. Perez, Am. J. Phys. {\bf 72}(2), 203 (2004).

\bibitem{phdCL}
C.L. Ching, Ph.D open Theses, https://scholarbank.nus.edu.sg/handle/10635/118204, (2014).

\bibitem{sakuraiqm} 
J.J Sakurai and J. Napolitano, Modern Quantum Mechanics (Cambridge University Press, UK, 2017).

\bibitem{antisnyder} 
C.L. Ching, C.X. Yeo, and W.K. Ng, Int. J. of Mod. Phys. A {\bf 32}, 1750009 (2017).


%Sect.(4)

\bibitem{pathria} 
R.K. Pathria, Statistical Mechanics (Butterworth Heinemann, UK, 1996).


%sect.(5)
\bibitem{bosso}
P. Bosso, S. Das and R.B. Mann, Phys. Rev. D {\bf 96}, 066008 (2017).
% deformed osscillator


%sect.(6)
\bibitem{kenfack}
A. Kenfack and K. Zyczkowski, J. Opt B: Quantum Semiclass. Opt. {\bf 6}, 396 (2004).
% wigner function

\bibitem{case} 
W.B. Case, Am. J. Phys. {\bf 76 (10)}, 937 (2008).

\bibitem{bassi} 
A. Bassi, A. Grobardt and H. Ulbricht, Class. Quantum Grav. {\bf 34}, 193002 (2017).
% gravitational decoherence

\bibitem{pikovskizych} 
I. Pokovski, M. Zych, F. Costa and C. Brukner, Nature Physics {\bf 11}, 668 (2015).
%universal decoherence due to gravitational time dilation

\end{thebibliography}
\end{document}